\documentclass[useAMS,usenatbib,manuscript,referee]{mn2e}
\usepackage{graphicx}
\usepackage{amsmath} 
\usepackage{amssymb}
\usepackage{natbib}

\newcommand{\beqa}{\begin{eqnarray}}
\newcommand{\eeqa}{\end{eqnarray}}

\def\hMpc{ \ifmmode{h^{-1}{\rm Mpc}}\else{$h^{-1}{\rm Mpc}$}\fi}
\def\hkpc{ \ifmmode{h^{-1}{\rm kpc}}\else{$h^{-1}{\rm kpc}$}\fi}
\def\hMsun{\ifmmode{h^{-1}M_\odot}  \else{$h^{-1}M_\odot$}\fi}

\def\up{_{\rm u}}
\def\down{_{\rm d}}

\newcommand{\Eq}[1]{Eq.\,(\ref{#1})}

\newcommand{\Fig}[1]{Fig.\,\ref{#1}}
\newcommand{\Sec}[1]{Sec.\,\ref{#1}}

\title[Radio signature from structure formation]
      {Radio signature of \\ cosmological structure formation shocks}

\author[Hoeft \&  Br\"uggen ]
    {
    Matthias Hoeft$^\ast$   and   Marcus Br\"uggen$^\ast$ \\
    International University Bremen, Campus Ring 1, 28759 Bremen, Germany \\
    }
\date{}

\def\LaTeX{L\kern-.36em\raise.3ex\hbox{a}\kern-.15em
    T\kern-.1667em\lower.7ex\hbox{E}\kern-.125emX}

\begin{document}

\label{firstpage}

\maketitle

\footnote[0]{$\ast$ e-mail: m.hoeft@iu-bremen.de, m.brueggen@iu-bremen.de}

\begin{abstract}

In the course of the formation of cosmological structures, large shock
 waves are generated in the intra-cluster medium. In analogy to
 processes in supernova remnants, these shock waves may generate a
 significant population of relativistic electrons which, in turn,
 produce observable synchrotron emission. The extended radio relics
 found at the periphery of several clusters and possibly also a
 fraction of radio halo emission may have this origin.  Here we derive
 an analytic expression for (i) the total radio power in the
 downstream region of a cosmological shock wave and (ii) the width of
 the radio-emitting region.  These expressions predict a spectral
 slope close to -1 for strong shocks. Moderate shocks, such as those
 produced in mergers between clusters of galaxies, lead to a somewhat
 steeper spectrum. Moreover, we predict an upper limit for the radio
 power of cosmological shocks.  Comparing our results to the radio
 relics in Abell\,115, 2256, and 3667, we conclude that the magnetic
 field in these relics is typically at a level of $0.1\,{\rm\mu G}$.
 Magnetic fields in the intra-cluster medium are presumably generated by the
 shocks themselves, this allows us to calculate the radio emission as a function of
 the cluster temperature. The resulting emissions agree very well with
 the radio power-temperature relation found for cluster
 halos. Finally, we show that cosmic accretion shocks generate less
 radio emission than merger shock waves. The latter may, however, be
 detected with upcoming radio telescopes.
 
\end{abstract}

\begin{keywords}
galaxies: clusters: general --
radio continuum: general --
diffuse radiation --
radiation mechanisms: non-thermal --
methods: analytical
\end{keywords}


\section{Introduction}

\label{sec-intro}

Diffuse radio objects that extend over several Mpc but have no optical counterpart have been observed in several clusters of galaxies. These objects fall into two categories: radio halos are centrally located in the cluster, they are fairly regular in shape and show very little polarisation. In contrast, radio relics appear at the periphery of clusters, have a filamentary morphology and are often highly polarised. Abell\,3667, for instance, shows a spectacular double relic \citep{roettgering:97}, where the two radio objects, separated by $\sim 4\:{\rm Mpc}$, straddle symmetrically the cluster X-ray emission. Similarly extended and remote relics have also been found in A115 \citep{govoni:01}, A2256 \citep{giovannini:99}, and A2345 \citep{giovannini:99} to name just a few. All of these clusters also show signs of an ongoing or recent merger: A115 has a double peak in the centre in X-ray observations \citep{shibata:99}, A2256 shows also a double peak \citep{sun:02}, A2345 is ass
 umed to be a dynamically young system
  since it shows multiple X-ray substructures \citep[see discussion in][]{dahle:02}, and in A3667, among other indications, a cold front has been found which is presumably related to a slightly supersonically moving substructure \citep{vikhlinin:02}. In particular, the last example suggests that the relics are located at the shock wave that has been produced by the substructure. \citet{roettiger:99} were able to reproduce the X-ray morphology and the location of the radio emission for this particular cluster in a merger simulation.

Even if the X-ray emission at the periphery of galaxy clusters is too
faint to identify clearly shock fronts, radio relics seem to trace
cosmological shock waves. Two models have been proposed for the
origin of radio relics: They could either be caused by diffusive shock
acceleration of electrons at the shock waves themselves
\citep{ensslin:98,miniati:01}, or they could be old radio bubbles that
are compressed by the passing shock wave and thus induced to emit observable
synchrotron emission again \citep{ensslin:01,ensslin:02,hoeft:04}. The
former relics are sometimes called {\it radio gischt}, while the
latter ones are known as {\it radio phoenix} \citep{kempner:04}. The
radio relic in A85 \citep{slee:01} may serve due to its size,
morphology, and strong polarization as prototype for the radio phoenix
class.

Radio halos, in contrast, do not directly trace the shock
fronts. Instead, the radio morphology is similar to the overall X-ray
morphology of the cluster \citep{govoni:01b}. The radio spectrum of
halos tends to steepen with radius
\citep{feretti:04,feretti:05b}. This may be regarded as evidence that
the emitting electrons are accelerated by turbulence in the ICM
\citep{brunetti:04}, which is a result of a
merger. However, the origin of the halos is still unclear. There is
increasing evidence that halos are indeed related to cluster mergers,
but it is a subject of a debate whether the relativistic electron
population that is responsible for the radio emission is emitted by a
central AGN, whether it is produced by strong shocks, whether it stems
from the decay of relativistic protons or whether it is generated by
turbulence in the downstream region of the shocks \citep[for a review
see e.\,g.][]{feretti:05}. Interestingly, the radio power of halos
scales with the cluster X-ray luminosity \citep{feretti:05} or, 
equivalently, with its temperature \citep{keshet:04}. Therefore, the 
non-thermal component in a galaxy cluster, if existent, seems to be 
related to the thermal plasma.

Numerical simulations show that the formation of the large-scale
structures leads to a variety of shocks in the intra-cluster medium
(ICM) and inter-galactic medium (IGM) \citep{miniati:00, ryu:03,
pfrommer:06, hoeft:06}. These shocks are crucial for the thermal state
of the ICM. As matter falls into the deep potential wells of galaxy
clusters, its shocks to the high temperature of the ICM of $\sim
10^7-10^8\:{\rm K}$.  \citet{ryu:03} distinguish between `internal' and
`external' shocks. The former affect material that is already heated
to roughly the cluster temperature, whereas the latter heat rather
cold gas for the first time. Hence, internal shocks are typically weak
while external shocks are very strong.

Cosmic rays with energies up to $\sim10^{15}\:{\rm eV}$ are presumably
accelerated at the shock fronts of supernova remnants
\citep[e.\,g.][]{berezhko:03,vink:03}. The acceleration is believed to
be caused by diffuse shock acceleration: Particles may be scattered
repeatedly between the upstream and the downstream regions, separated
by the shock. Each time the particle crosses the front, it gains
kinetic energy \citep[see for a
review][]{axford:78,bell:78a,bell:78b,blandford:78}. Even though the
physical conditions of shock waves in the ICM are somewhat different
from those in supernova remnants, one would also expect diffuse shock
acceleration to be efficient at cosmological shocks.

Various groups have tried to estimate the radio emission of the shock
fronts by combining the probability distribution of cosmic shock
fronts with models for diffuse shock acceleration and the subsequent
synchrotron emission. \citet{keshet:04} found that 10\,\% of the
extragalactic radio background below 500\,MHz should be caused by
cosmic structure formation shocks for $\Lambda$CDM cosmology. They assumed 
that the emission comes mainly from external accretion shocks, in
particular those that surround massive clusters of galaxies. Essential
for the radio emission is the strength of the magnetic field since the
emission is synchrotron emission. Unfortunately, the magnetic field
distribution in clusters of galaxies and their surroundings is
difficult to determine and still poorly constrained
\citep[for a review see][]{govoni:04}.

In this paper, we compute the radio emission in the downstream region
 of a shock front. We assume that diffuse shock acceleration produces
 a relativistic electron population with a power-law distribution in
 the energy spectrum (\Sec{sec-DSA}). As the plasma moves downstream,
 the high-energy electrons cool by inverse Compton and synchrotron
 losses (\Sec{sec-cool}). The main objective of this paper is to
 derive an analytic expression for (i) the total radio power in the
 downstream region of a cosmological shock wave and (ii) the width of
 the radio-emitting region. In particular, we wish to explore the
 dependence of the power, spectral shape and spatial extent of the
 radio emission on the magnetic field and the shock parameters.

We apply our model to the relics observed
 in Abell\,115, 2256, and 3667 and infer the magnetic field strength
 in the relic region (\Sec{sec-examples}). Assuming that the magnetic
 field energy density is a fixed fraction of the thermal energy allows
 us to express the radio power as a function of the cluster temperature
 (\Sec{sec-P-T}). Finally, we make predictions for the radio emission
 of cosmic accretion shocks (\Sec{sec-cosmic}).

\section{Radio emission by non-radiative shocks}

\subsection{Non-radiative shocks}
\label{sec-shock}

A shock surface separates two regions: The upstream plasma moves with
velocity $v\up$ towards the shock front, while the downstream plasma
departs with $v\down$. As the plasma passes through the shock front,
mass, momentum, and energy fluxes are conserved, which is expressed in
the Rankine-Hugoniot relations:
\begin{eqnarray}
  \rho\up v\up
  &=&
  \rho\down v\down
  \nonumber
  \\
  P\up +  \rho\up v\up^2
  &=&
  P\down +  \rho\down v\down^2
  \\
  \frac{1}{2}v\up^2 + u\up + \frac{P\up}{\rho\up}
  &=&
  \frac{1}{2}v\down^2 + u\down + \frac{P\down}{\rho\down}
  \nonumber
  ,
  \label{eq-conservation}
\end{eqnarray}
where $\rho$ denotes mass density, $P$ pressure, and $u$ specific
internal energy. On timescales relevant for the shock propagation, we assume that the intra-cluster medium behaves like a polytropic, ideal gas. Hence, the pressure can be written as
\begin{eqnarray}
  P
  =
  (\gamma - 1) \,
  \rho \,
  u
  ,
  \label{eq-pressure-poly}
\end{eqnarray}
where $\gamma$ is the adiabatic index, and the specific internal energy depends only on the gas temperature, $T$,
\begin{eqnarray}
  (\gamma-1) \mu m_{\rm p} \:
  u
  =
  k_{\rm B} T
  ,
  \label{eq-rel-u-T}
\end{eqnarray}
where $m_{\rm p}$ is the proton mass and $\mu$ is the molecular weight. We determine the latter by adopting a fully ionised plasma with primordial chemical composition. In contrast to the conserved properties, the entropy of the plasma is increased by the dissipation at the shock front. For simplicity the term `entropy' will refer throughout this paper to a monotonic function of it, namely the entropic index,
\begin{eqnarray}
  S 
  \equiv
  u \,
  \rho^{1-\gamma} 
  .
  \label{eq-def-entr-index}
\end{eqnarray}

The strength of non-radiative shocks in a polytropic gas can be characterised by a single parameter, e.\,g. the compression ratio 
\begin{eqnarray}
  r 
  \equiv
  \frac{\rho\down}{\rho\up}
  ,
  \label{eq-def-compress-ratio}
\end{eqnarray}
or, equivalently, by the entropy ratio
\begin{eqnarray}
  q 
  \equiv
  \frac{S\down}{S\up}
  .
  \label{eq-def-entr-ratio}
\end{eqnarray}
The ratio of the specific internal energies can be given as a function of the compression and entropy ratios
\begin{eqnarray}
  \frac{u\down}{u\up}
  =
  \frac{S\down}{S\up} \,
  \left(
  \frac{\rho\down}{\rho\up}
  \right)^{\gamma-1}
  =
  q \,
  r^{\gamma-1} 
  .
  \label{eq-u-ratio}
\end{eqnarray}
Combining the conservation laws, \Eq{eq-conservation}, and using the definition of the entropy, \Eq{eq-def-entr-index}, allows us to relate the compression and the entropy ratios by the implicit equation
\begin{eqnarray}
  r
  =
  \frac{ (\gamma+1) \, q \, r^\gamma + (\gamma-1)}
       { (\gamma-1) \, q \, r^\gamma + (\gamma+1)}
  .
  \label{eq-r-from-q}
\end{eqnarray}
Hence, if we can determine the entropy ratio, $q$, for a shock we can compute the compression ratio by \Eq{eq-r-from-q} and finally the ratio of internal energies by \Eq{eq-u-ratio}.

\begin{figure}
  \begin{center}
  \includegraphics[width=0.7\textwidth,angle=-90]{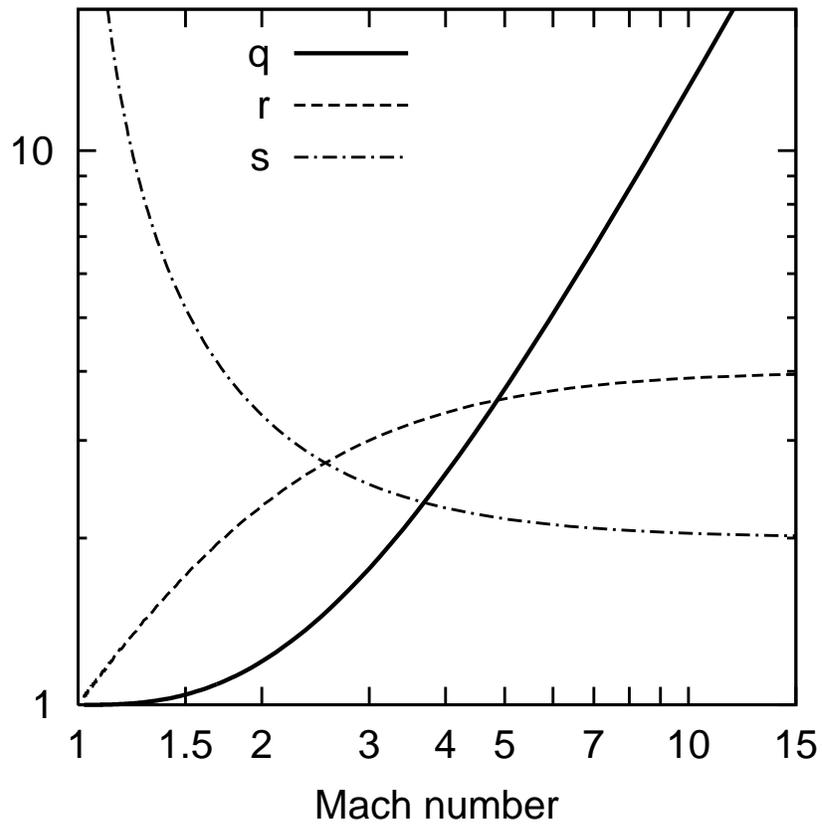}
  \end{center}
  \caption
  { 
Entropy ratio, $q=S\down/S\up$, compression ratio, $r=\rho\down/\rho\up$, and spectral index, $s$, of the electron spectrum generated by diffuse shock acceleration as a function of the Mach number. 
  }
  \label{fig-Msqr}
\end{figure}

Customarily, the strength of a shock is characterised by the upstream Mach number
\begin{eqnarray}
  {\cal M}
  =
  \frac{v\up}{c\up}
  ,
  \label{eq-def-mach}
\end{eqnarray}
where $c\up$ denotes the upstream sound speed, which depends on the specific internal energy by
\begin{eqnarray}
  c\up^2
  =
  \gamma
  (\gamma-1) \,
  u\up
  .
  \label{eq-sound-speed}
\end{eqnarray}
Invoking flux conservation, \Eq{eq-conservation}, we can rewrite the Mach number
\begin{eqnarray}
  {\cal M}^2
  =
  \frac{1}{c\up^2} \:
  \frac{\rho\down}{\rho\up} \:
  \frac{P\down - P\up}{\rho\down - \rho\up}
  =
  \frac{r}{\gamma} \: 
  \frac{ q r^\gamma - 1 }{ r - 1 }
  .
  \label{eq-mach-from-q-r}
\end{eqnarray}
For large Mach numbers the entropy ratio is proportional to ${\cal M}^2$, see \Fig{fig-Msqr}. For later use we rewrite the downstream velocity 
\begin{eqnarray}
  v\down^2
  =
  \frac{\gamma - 1}{q r^\gamma } \:
  \frac{qr^\gamma - 1}{r-1} \;
  u\down
  \equiv
  C_v^2 \:
  u\down
  ,
  \label{eq-v-d}
\end{eqnarray}

\subsection{Diffuse shock acceleration}

\label{sec-DSA}

In supernova remnants, there is evidence that electrons and protons
are accelerated by diffuse shock acceleration (DSA) to energies of
$\rm\sim10^{15}\,eV$ \citep[e.g.][]{berezhko:03,vink:03}. In DSA,
particles are accelerated by multiple shock crossings, in a
first-order Fermi process. If the shock thickness is much smaller than
the diffusion scale which in turn has to be much smaller than the
curvature of the shock front a one-dimensional diffusion-convection
equation can be solved
\citep{axford:78,bell:78a,bell:78b,blandford:78}. The result is that the energy spectrum of suprathermal
electrons is a power-law distribution,
$n_E \propto E^{-s}$. The spectral index, $s$, of the accelerated particles is only related to
the compression ratio at the shock front
\begin{eqnarray}
  s
  =
  \frac{r+2}{r-1}
  .
 \label{eq-s-from-r}  
\end{eqnarray}
For strong, non-radiative shocks with Mach number $\gtrsim$10, the compression ratio is always close to 4, hence the slope $s$ is always close to 2. This allows to explain the spectrum of cosmic rays over a huge range of energies and may be considered as a piece of evidence for DSA. For a review of diffuse shock acceleration see \citet{drury:83,blandford:87,jones:91,malkov:01}.

Diffuse shock acceleration theory suffers from the complexity of real
shock fronts and the highly non-linear interaction between different
processes such as cosmic rays and magnetic waves. The efficiency of
accelerating electrons and protons has been inferred from the
observation of supernova remnants. For instance, \citet{dyer:01}
estimated for SN\,1006 that a few percent of the shock energy is
transferred to supra-thermal particles. Similar to \citet{keshet:04},
we assume that a fixed fraction $\xi_e$ of thermal energy injected at
the shock front goes into the acceleration of suprathermal electrons.

Moreover, we multiply the spectrum by an upper
cut-off factor since electrons can only be accelerated to finite
energies. Typically, the maximum energy is estimated by comparing the
acceleration $e$-folding time and the cooling time \citep{keshet:03}.
We use here a smoothed high-energy cut-off, the reason for this
particular choice will be become clear in the next section. Thus, the
electron spectrum generated by DSA, is given by 
\begin{eqnarray}
  n_E(E)
  \equiv
  \frac{{\rm d}n_{\rm e}}
       {{\rm d}E}
  =
  \left\{
  \begin{array}{r@{\quad:\quad}l}
    n_{\rm e} \,
    C_{\rm spec} \:
    \frac{ 1 }{ m_{\rm e} c^2} \,
    \tilde{e}^{-s}
    \left\{
      1 
      - 
     \frac{ \tilde{e} }
          { \tilde{e}_{\rm max} }
    \right\}^{s-2}
    &
    \tilde{e} < \tilde{e}_{\rm max}
    \\
    0
    &
    {\rm elsewhere}
    \\
  \end{array}
  \right.
  ,
  \label{eq-e-supr-spec}  
\end{eqnarray}
where we have used the abbreviation $\tilde{e} = E/m_{\rm e}c^2$. The normalisation constant can be interpreted as follows: We compute the electron number density of electrons with energies in the interval $[E,E+\Delta E]$. If we choose $E=\Delta E =  m_{\rm e}c^2$, the number density becomes
\begin{eqnarray}
  n_E \Delta E
  = 
  n_{\rm e} C_{\rm spec}
  .
  \nonumber
\end{eqnarray}
Hence, $C_{\rm spec}$ gives basically the fraction of electrons at $E=m_{\rm e}c^2$.

We determine the fraction of suprathermal electrons by postulating that a
fixed fraction $\xi_{\rm e}$ of the energy injected at the shock front
goes into the acceleration of electrons to suprathermal energies. After
passing the shock, the downstream gas has gained a thermal energy of
\begin{eqnarray}
  u\down\rho\down - r^{\gamma}u\up\rho\up
  =
  u\down\rho\down
  \frac{q-1}{q}
  ,
  \nonumber
\end{eqnarray}
where $r^{\gamma}u\up\rho\up$ would be the energy density after pure adiabatic compression. Hence, the spectrum can be normalised by
\begin{eqnarray}
  \int\nolimits_{E_{\rm min}}^\infty \: {\rm d}E \;
  n_E(E) \;
  E
  =
  \xi_e \,
  u\down\rho\down
  \frac{q-1}{q}
  . 
  \label{eq-normalize-e-spec-energy}
\end{eqnarray}
A crucial parameter is the minimum energy, $E_{\rm min}$, above which electrons are considered to be suprathermal. It is  particularly important for spectra significantly steeper than $s=2$ since for these spectra most of the suprathermal energy is carried by electrons immediately above $E_{\rm min}$. Therefore, the normalisation of the spectrum depends on $E_{\rm min}$. Hybrid simulation of a collisionless shock indicate that there is continuous transition from the thermal to the suprathermal distribution \citep{bennett:95}. Simulations of electron acceleration at high Mach number shocks may rely on injection of electrons with a fixed energy. However, the typical injection energy corresponds to thermal energy in the ICM \citep{levinson:94}, therefore, the suprathermal spectrum is expected to be a continuous extension of the thermal spectrum.  
We assume here that the thermal Maxwell-Boltzmann distribution goes
over continuously into the power-law spectrum of the suprathermal electrons
\begin{eqnarray}
  n_E^{\rm Maxwell}(E_{\rm min})
  =
  n_E(E_{\rm min})
  .
  \nonumber
\end{eqnarray}
This condition leads to an implicit equation for $E_{\rm min}$ that has to be solved simultaneously with the normalisation of the spectrum. Basically, the transition energy is tightly coupled to the temperature of the plasma,
\begin{eqnarray}
  E_{\rm min}
  \sim
  10
  \times
  k_{\rm B}T
  ,
  \nonumber
\end{eqnarray}
if no extreme parameters are chosen. 

We can write for the normalisation of the electron spectrum    
\begin{eqnarray}
  \xi_e \,
  u\down\rho\down
  \frac{q-1}{q}
  =
  n_{\rm e} \,
  C_{\rm spec} \,
  m_{\rm e} c^2 \,
  \int_{\tilde{e}_{\rm min}}^\infty {\rm d}\tilde{e} \;
  \tilde{e}^{1-s} \,
  \left\{
   1 
   -
   \frac{\tilde{e}}
        {\tilde{e}_{\rm max}}
   \right\}^{s-2} 
   \equiv
   n_{\rm e} \,
   C_{\rm spec} \,
   m_{\rm e} c^2 \,
   I_{\rm spec}
   ,
  \nonumber
\end{eqnarray}
where

\begin{eqnarray} 
I_{\rm spec} = 
\int_{\tilde{e}_{\rm min}}^\infty {\rm d}\tilde{e} \;
  \tilde{e}^{1-s} \,
  \left\{
   1 
   -
   \frac{\tilde{e}}
        {\tilde{e}_{\rm max}}
   \right\}^{s-2}
   .
\end{eqnarray}

\begin{figure}
  \begin{center}
  \includegraphics[width=0.7\textwidth,angle=0]{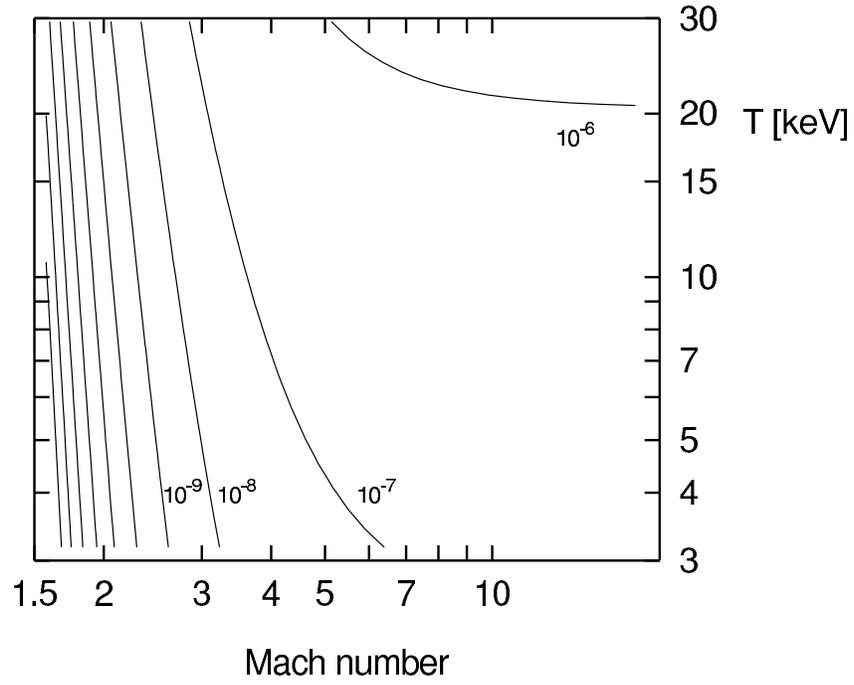}
  \end{center}
  \caption
  { 
Contours of the fraction of relativistic electrons with $\gamma_{\rm L} > 300$. 
Contour levels are at $10^{-6}$, $10^{-7}$, $10^{-8}$, etc.. The energy fraction of suprathermal electrons is $\xi_{\rm elec} = 0.05$. The initial cutoff in the spectrum is at $\tilde{e} = 10^{7}$.
  }
  \label{fig-nFrac}
\end{figure}

Hence, the constant becomes
\begin{eqnarray} 
  C_{\rm spec}
  =
  \underbrace{
    \xi_e
    \frac{u\down}{c^2}
    \frac{m_{\rm p}}{m_{\rm e}}}_{C_{\rm spec}^p} \;
  \underbrace{
    \frac{(q-1)}{q}
    \frac{1}{I_{\rm spec}}
  }_{C_{\rm spec}^q}
  ,
  \label{eq-C1}
\end{eqnarray}
where the first factor, $C_{\rm spec}^p$, includes the plasma
contributions, while the second factor, $C_{\rm spec}^q$, incorporates
the dependencies from the shock strength. Since the electron spectrum is steep
for Mach numbers $\lesssim 2$, the fraction of relativistic electrons is low
in this regime, see \Fig{fig-nFrac}. Note that also the temperature of the
downstream plasma affects the relativistic electron density. For Mach numbers
$\gtrsim 10$ the fraction of relativistic electrons is virtually constant
since the electron spectrum varies only negligibly.\\

  In this paper, we only consider the acceleration of {\it electrons} in order
  to compute the radio emission from cosmological shocks.  Clearly, there is
  evidence that also protons are accelerated to highly relativistic energies
  at shocks, probably even more efficiently than electrons, c.f.
  \citet{levinson:94}. Because of their larger mass, protons cause
  significantly less synchrotron emission than electrons.  However, in
  inelastic collisions, relativistic protons produce electron/positron pairs
  which, in turn, radiate synchrotron emission.  In the ICM, though, the life
  time of relativistic protons is about
  the Hubble time \citep{dennison:80}. As a result, relativistic protons gradually
  accumulate in the cluster potential. While the emission by secondary
  electrons may contaminate to some degree the emission by primary electrons, 
  it is unlikely to account for the bulk of the radio
  emission from radio relics.  Radio relics have a filamentary morphology and
  the emission is confined to regions close to the shock. This morphology is
  difficult to explain with superthermal protons and is more likely caused by
  short-lived, primary electrons. The emission from secondary electrons and
  positrons produced by relativistic protons may be responsible for the
  cluster-wide emission observed in radio halos. However, acceleration of
  protons has another consequence.  Suprathermal protons can affect the
  pressure in the gas and thus affect the structure of the shock. An admixture
  of relativistic protons lowers the adiabatic exponent, $\gamma$, to below
  the value for a non-relativistic gas of $5/3$. In the extreme case of a
  fully relativistic gas, $\gamma$ is $4/3$.  An effect of a lower $\gamma$ is
  that the compression ratio of the shock can be higher, and, as a result,
  \Eq{eq-s-from-r} becomes inaccurate.  Moreover, the back-reaction of the
  relativistic particles on the shock leads to non-linear particle
  acceleration, which results in flatter particle spectra than the power-laws
  obtained in the test-particle approximation.  \citet{amato:05} have
  demonstrated the effects of non-linear particle acceleration on shock
  structure and particle spectra. While there is some evidence for non-linear
  effects in supernova remnants, the details of this process remain
  controversial \citep{ellison:05}. In particular, it is unclear to what
  extent non-linear shock acceleration takes place in cosmological shocks.
  Here we will adopt the results obtained with the linear test-particle
  approximation.

\subsection{Radiative cooling of high energy electrons}
\label{sec-cool}

High-energy electrons in the ICM cool mainly by inverse Compton scattering (IC) with CMB photons and by synchrotron radiation losses
\begin{eqnarray}
   \frac{{\rm d} E}{{\rm d} t}
  &=&
  - \frac{C_{\rm cool}}{m_{\rm e}c^2}  E^2
  ,
  \label{eq-electron-cooling} 
\end{eqnarray}
where $E$ is the energy of an electron. The cooling constant is given by
\begin{eqnarray}
  C_{\rm cool}
  \equiv
  \frac{4 \, \sigma_{\rm T}}{3 \, m_{\rm e} \, c} \, \{ u_{\rm CMB} + u_B \} 
  ,
  \label{eq-cooling-constant} 
\end{eqnarray}
where $\sigma_{\rm T}$ is the Thomson cross section, $u_{\rm CMB}$ and
$u_B$ are the energy densities of the radiation background and the
magnetic field, respectively. We have neglected the effects of
adiabatic cooling and heating which depend on the local flow
conditions. Adiabatic cooling can in principle be included but for our
purpose we will ignore it here. Electron number conservation leads to a PDE for the energy spectrum 
\begin{eqnarray}
  \frac{ \partial n_E} {\partial t}
  =
  C_{\rm cool} \tilde{e}^2 \,
  \frac{ \partial n_E} {\partial \tilde{e}}
  +
  2 C_{\rm cool} \tilde{e} \,
  n_E
  ,
  \label{eq-PDE-spectrum} 
\end{eqnarray}
with the solution \citep{kardashev:62}
\begin{eqnarray}
  n_E(E,t)
  =
  \left\{
    \begin{array}{c@{\quad:\quad}l}
        n_{\rm e} \,
        C_{\rm spec} \,
        \frac{1}{m_{\rm e}c^2} \:
        \tilde{e}^{-s}
        \left\{ 
                1 
                - 
                \left(
                    \frac{1}{\tilde{e}_{\rm max}}
                    +
                    C_{\rm cool} t 
                \right)
                \tilde{e}       
        \right\}^{s-2}  &  \tilde{e} \, C_{\rm cool} \, t < 1 - \tilde{e}/\tilde{e}_{\rm max} \\
                0                        &  {\rm elsewhere }
    \end{array}
  \right.
  .
  \label{eq-e-spectrum-solution} 
\end{eqnarray}
We have chosen the cut-off factor in \Eq{eq-e-supr-spec} consistently with this solution, hence the initial spectrum is already a solution to \Eq{eq-PDE-spectrum}.

The energy densities of the cosmic radiation background with temperature $T_{\rm CMB} = T_{\rm CMB,0}(z+1)$ and of the downstream magnetic field, $B\down$ are
\begin{eqnarray}
  u_{\rm CMB} 
  &=& 
  a_{\rm rad} \, T_{\rm CMB, 0}^4 \, (z+1)^4
  \equiv
  \frac{B_{\rm CMB}^2}{8 \pi}
  \label{eq-def-u-CMB}
  \\ 
  u_B 
  &=& 
  \frac{B\down^2}{8 \pi}
  ,
  \label{eq-def-u-B}
\end{eqnarray}
respectively, where $z$ is the redshift and $a_{\rm rad}$ the
radiation constant. For convenience, we have introduced a magnetic
field equivalent for the background radiation, $B_{\rm CMB}$. For
$z=0$, $B_{\rm CMB}$ is about $3.24\:{\rm\mu G}$. Since we are
primarily interested in radio emission which is produced by
relativistic electrons, it is convenient to
express the related equations in terms of the Lorentz factors,
$\gamma_{\rm L}$, of the electrons. Only electrons with $\gamma_{\rm
L} \gg 1$ contribute significantly to the radio emission and we can approximate the Lorentz factor 
\begin{eqnarray}
  \gamma_{\rm L}
  =
  \frac{E}{m_e c^2}
  + 
  1
  \approx
  \frac{E}{m_e c^2}
  =
  \tilde{e}
  \qquad
  {\rm for}
  \qquad
  E 
  \gg
  m_e c^2
  .
  \label{eq-def-gamma}
\end{eqnarray}
For any observing frequency, there is a minimum Lorentz factor below
which the radio emission is negligible. On the other hand,
\Eq{eq-e-spectrum-solution} shows that the electron spectrum has a
time-dependent, maximum energy. For any Lorentz factor we can
determine the time when then maximum energy in the spectrum equals the
given $\gamma_{\rm L}$, which is given by
\begin{eqnarray}
  t_{\rm cool}(\gamma_{\rm L})
  =
  \frac{1}{C_{\rm cool} \,E(\gamma_{\rm L})}
  =
  2.32 \times 10^{12} \, \gamma_{\rm L}^{-1} \, (z+1)^{-4} \, {\rm yr}
  ,
  \nonumber
\end{eqnarray}
where the right expression is computed for pure inverse-Compton
cooling (we have assumed here $\tilde{e}/\tilde{e}_{\rm
max} \ll 1$). After a time $t_{\rm cool}(\gamma_{\rm L})$ there are no electrons with Lorentz factor above $\gamma_{\rm L}$.

\subsection{The synchrotron emission}
\label{sec-sync}

The synchrotron power of a single electron, $P_{\rm e}$, that moves with a Lorentz factor $\gamma_{\rm L}$ in a magnetic field $B$ is \citep[for an introduction see][]{rybicki:86}
\begin{eqnarray}  
  \frac{ {\rm d} P_{\rm e}(\gamma_{\rm L}, \nu_{\rm obs}) }{ {\rm d}\nu }
  =
  \frac{ \sqrt{3} \, e^3 B \sin \alpha } { m_{\rm e} c^2 }
  F \left( \frac{ \nu_{\rm obs} }{ \nu_{\rm c} } \right) 
  \,\equiv\,
  C_{\rm sync} \,
  F \left( \frac{ \nu_{\rm obs} }{ \nu_{\rm c} } \right) 
  .
  \label{eq-synchro-basic}
\end{eqnarray}
where $e$ is the electron charge, $\alpha$ the angle enclosed between the direction of the motion and the magnetic field, and $\nu_{\rm c}$ is the characteristic frequency
\begin{eqnarray}  
  \nu_{\rm c} 
  =
  \frac{ 3 \gamma_{\rm L}^2 e B \sin \alpha } { 4 \pi \, m_{\rm e} c }
  .
  \nonumber
\end{eqnarray}
In the following, we assume  that the directions of the electron
motion and the magnetic field are uncorrelated, therefore we take the
average value $\langle \sin \alpha \rangle = \pi/4$, where the average
is over solid angle.
The function $F(x)$ has to be integrated numerically,
\begin{eqnarray}  
  F(x)
  =
  x \,
  \int\nolimits_x^\infty {\rm d}\xi 
  \: 
  K_{\frac{5}{3}} (\xi) 
  ,
  \nonumber
\end{eqnarray}
where $K$ is the modified Bessel function.
The emission per volume is given by the convolution
\begin{eqnarray}  
  \frac{{\rm d}^2 \, P (\nu_{\rm obs})} {{\rm d} V \: {\rm d}\nu }
  =
  \int_0^\infty  {\rm d } E \,
  n_E \:
  P_{\rm e}
  = 
  C_{\rm sync} \int_0^\infty  {\rm d }\gamma_{\rm L} \,
  m_{\rm e}c^2 \;
  n_E( \gamma_{\rm L},t) \:
  F(\gamma_{\rm L},\nu_{\rm obs}) \ .
  \nonumber
\end{eqnarray}
We introduce a new variable  
\begin{eqnarray}  
  \tau
  \equiv 
  \sqrt{\frac{ 3 \,e B } {  \nu_{\rm obs} \, 16 \,m_{\rm e} c }}  \;
  \gamma_{\rm L}
  \equiv
  C_\tau \, 
  \gamma_{\rm L}
  =
  4.85 \times 10^{-5} \:
  \left\{
    \frac{B}{\rm  \mu G}
  \right\}^\frac{1}{2}
  \left\{
    \frac{\rm 1.4 \, GHz}{\nu_{\rm obs}}
  \right\}^\frac{1}{2} \:
  \gamma_{\rm L}
  .
  \label{eq-def-tau}
\end{eqnarray}
Substituting $\gamma_{\rm L}$ by $\tau$, the integration becomes
\begin{eqnarray}  
  \frac{{\rm d}^2 P (\nu_{\rm obs})} {{\rm d} V \: {\rm d}\nu }
  =
  \frac{C_{\rm sync} \,  m_{\rm e} c^2 }{C_\tau} \:
  \int_0^\infty  {\rm d }\tau \:
  n_E(\tau,t) \,
  F\left( \frac{1}{\tau^2} \right)
  .
  \label{eq-cum-synchro}
\end{eqnarray}
Assuming a power-law spectrum for $n_E$, we compute the cumulative emission, see \Fig{fig-cum-synchro}. The bulk of the emission comes from $\tau_{\rm half} \sim 0.03$. This implies that electrons with Lorentz factors about
\begin{eqnarray}  
  \gamma_{\rm L, half}
  =
  6.2 \times 10^2 \:
  \left\{
    \frac{\rm \mu G}{B}
  \right\}^\frac{1}{2}
  \left\{
    \frac{\nu_{\rm obs}}{\rm 1.4 \, GHz}
  \right\}^\frac{1}{2} \:
  \label{eq-gamma-half}
\end{eqnarray}
contribute most to the synchrotron emission. Note that at
$\gamma_{\rm L} \sim 3 \times 10^2$ and below, Coulomb losses
constitute the most important cooling mechanism for suprathermal
electrons, see e.g. \citet{sarazin:99}. Hence, at very low observing
frequencies in regions with a rather strong magnetic field, Coulomb
cooling becomes important. For instance, {\sc Lofar} may observe at
30\,MHz a radio halo region where the magnetic field is of the order
of $3\:{\rm\mu G}$. \Eq{eq-gamma-half} indicates that the bulk of 
the radio emission would come from electrons with Lorentz factor 
$\gamma_{\rm L} \sim 50$. For Lorentz factor that low Coulomb cooling
is important, hence, our analysis may overestimate the synchrotron 
emission at very low observing frequencies.

\subsection{Total emission behind the shock front}
\label{sec-total}

In \Sec{sec-DSA}, we worked out the relativistic spectrum of
 electrons accelerated at the shock front, in \Sec{sec-cool} we
 worked out the evolution of the spectrum subject to cooling, and in
 \Sec{sec-sync} we computed the synchrotron emissivity for a
 homogeneous electron population. Now we combine these results, to
 compute the total emission behind the shock front. We adopt the
 following scenario: Diffuse shock acceleration generates a population
 of suprathermal electrons. In energy space, these electrons are
 distributed initially according to a power-law. While they advect
 with the downstream plasma, the relativistic electrons cool by
 synchrotron and inverse Compton losses. With increasing distance to
 the shock front, the maximum electron energy decreases and the radio
 emission diminishes. The total emission is obtained by summing up all
 contributions from the plasma from the shock front to the distance
 where the electron spectrum is too cool to allow any further radio
 emission.

To compute the spatial distribution of the emission, we have to model
the downstream region of the shock front. Instabilities of 
waves propagating at the shock surface presumably originate turbulence 
in the downstream plasma immediately behind the shock surface. If this turbulence
can also be found at larger distances from the shock surface
depends strongly on the viscosity of the ICM, which is still rather uncertain
\citep[see e.\,g.][]{fabian:05}. If the viscosity is indeed close to
the Spitzer value of a hot plasma, as suggested by the H$\alpha$
filaments surrounding NGC\,1275, the formation of turbulence would be
significantly hampered. Hence, we assume the flow in the downstream
region to be steady and laminar.\\

  Diffusion can alter the spatial distribution of suprathermal electrons. In the
  absence of magnetic fields, the diffusion length is dominated by Coulomb
  collisions. The presence of even very weak magnetic fields suppresses
  diffusion in various ways. In directions perpendicular to magnetic field lines
  electrons propagate roughly with the Bohm diffusion rate. The resulting
  distance covered within the cooling time of even very energetic particles
  is negligible for our purposes. Parallel to the magnetic field lines the
  electrons propagate with the drift velocity, provided the gyroradius is
  significantly smaller than the coherence length of the magnetic field. The
  gyroradius of relativistic electrons is given by

\begin{equation}
  r_{\rm g}
  =
  \gamma_{\rm L} \,
  \frac{m_{\rm e}c^2}{eB} \approx 6\times10^{-8}\:{\rm kpc} \left (
  \frac{\gamma_{\rm L}}{10^4}\right ) \left ( \frac{B}{1\,\mu G}\right )^{-1}
  .
  \nonumber
\end{equation}

Hence, the gyroradius of suprathermal electrons in the ICM is of the order
  of $10^{12} - 10^{14}\:{\rm cm}$. The  correlation length of magnetic fields
  generated at the shock surface is expected to be of the order of
  $10^{10}\:{\rm cm}$ \citep{medvedev:06}, but it is also expected to increase
  rapidly \citep{medvedev:05}. As a result, electrons drift virtually freely
  along the field lines. Thus, diffusion along field lines is the most efficient
  transport, any tangling of field lines inside the gyroradius reduces the
  diffusion \citep{colafrancesco:98}.  The electron drift is hampered by pitch
  angle scattering with plasma waves, in particular with Alfv{\'e}n waves
  \citep{wentzel:74}. These waves are generated by fast electrons and protons
  that move with super-Alfv{\'e}n speed. Thus, collective effects in the 
  plasma suppress the drift
  parallel to the field lines. Even if the precise treatment of electron
  diffusion is impossible since the actual state of the downstream plasma is
  not very well known, the collective effects seem to restrict the diffusion of suprathermal
  electrons to much smaller scales than those that we discuss in
  the context of radio relics, see \Sec{sec-examples}. This allows us to assume
  that relativistic electrons purely advect with the plasma. Similar conclusions have been drawn in
  related studies that show, for example, that radio halos cannot be explained by the diffusion
  of electrons from a central radio galaxy \citep{jaffe:77,dennison:80,fujita:01}. 
 \\

We now compute the total emission from the downstream region. The total emission per shock surface area $A$ is given by the integral over the whole downstream region, the width of which is here parameterized by $y$ 
\begin{eqnarray}  
  \frac{{\rm d}^2 P ( \nu_{\rm obs} )}{{\rm d} A \: {\rm d} \nu}
  =
  \int\nolimits_0^\infty {\rm d } y \;
  \int_{E_{\rm min}}^\infty  {\rm d }E \;
  n_E (E,t)\,
  P(\gamma_{\rm L}, \omega_{\rm obs}) 
  .
  \label{eq-total-emission}
\end{eqnarray}
In any given volume element, the electron spectrum depends on the time
passed since the acceleration at the shock front due to cooling. If we
assume that the suprathermal electrons advect with the plasma at
constant speed, the distance to the shock front and the time passed since acceleration are related by $y=v\down t$. Hence, we can write
\begin{eqnarray}  
  C_{\rm cool} \, \tilde{e} \, t 
  =
  C_{\rm cool} \:
  \gamma_{\rm L} \:
  \frac{y}{v\down}
  =
  \frac{C_{\rm cool}}{C_\tau v\down} \:
  \tau 
  y
  .
  \nonumber
\end{eqnarray}
Using the following substitution
\begin{eqnarray}  
  \eta
  =
  \frac{C_{\rm cool}}{C_\tau v\down} \,
  y
  ,
  \label{eq-def-eta}
\end{eqnarray}
the integral \Eq{eq-total-emission} becomes
\begin{eqnarray}  
  \frac{{\rm d}^2 P ( \nu_{\rm obs} )}{{\rm d} A \: {\rm d} \nu}
  =
  C_{\rm all}
  \int\nolimits_0^\infty {\rm d } \eta \;
  \int_{\tau_{\rm min}}^\infty  {\rm d } \tau \,
  \tau^{-s} \:
  (1 - \tau \eta)^{s-2} \: 
  F\left( \frac{1}{\tau^2}\right) 
  \equiv
  C_{\rm all}
  I_\Psi(s)
  .
  \label{eq-C-all}
\end{eqnarray}
The lower limit, $\tau_{\rm min}$, is in principle defined by the energy where the energy spectrum goes from thermal to suprathermal. However, as discussed above, the bulk of the radiation comes from the region with $\tau \sim 0.03$, which is by orders of magnitude above the value of $\tau_{\rm min}$. Thus, we are free to use 0 as lower integration limit. The rapid decrease of $F(1/\tau^2\to\infty)$ ensures that the integrand is well-behaved. As a result, the integral depends only on the spectral index of the electron spectrum, $s$.

\begin{figure}
  \begin{center}
  \includegraphics[width=0.7\textwidth,angle=-90]{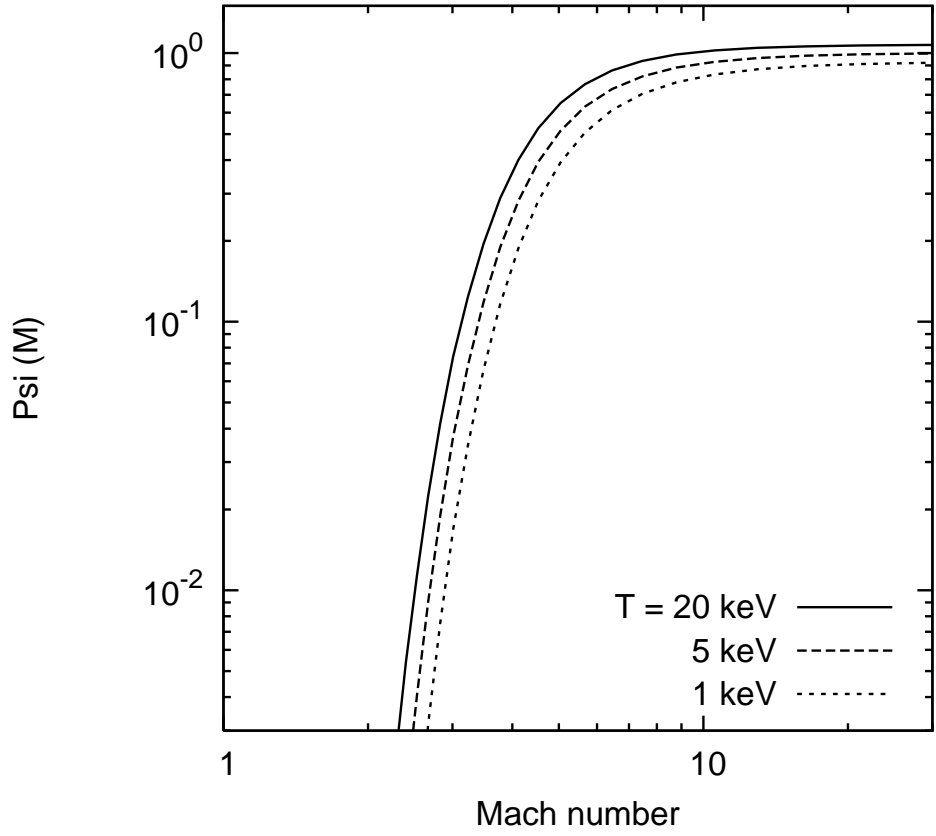}
  \end{center}
  \caption
  { 
$\Psi$ as function of the Mach number. Also the downstream temperature has a small impact.
  }
  \label{fig-psi}
\end{figure}

Substituting the normalisation of the electron spectrum, \Eq{eq-e-spectrum-solution}, the emission of a homogenous volume \Eq{eq-cum-synchro}, and \Eq{eq-def-eta} to the total emission \Eq{eq-C-all}, we find
\begin{eqnarray}  
  C_{\rm all} 
  & = &
  m_{\rm e}c^2 
  \frac{C_{\rm sync}}{C_\tau}
  \times
  n_{\rm e}
  C_{\rm spec}
  \frac{1}{m_e c^2}
  C_\tau^s
  \times
  \frac{C_\tau v\down}{C_{\rm cool}}
  =
  n_{\rm e} \,
  C_{\rm spec}  \,
  C_{\rm sync}\,
  C_\tau^s \:
  \frac{C_v \sqrt{u\down}}{C_{\rm cool}} 
  \nonumber
  ,
\end{eqnarray}
where we have also used $v\down = C_v \sqrt{u\down}$, see
\Eq{eq-v-d}. Now, it is useful to combine all factors that depend
on the strength of the shock in \Eq{eq-C-all} in
\begin{eqnarray}  
  \Psi({\cal M})
  & = &
  C_\Psi \:
  C_v ({\cal M}) \:
  \left( C_\tau( 1 \: {\rm \mu G}, 1.4 \: {\rm GHz }) \right)^s\:
  C_{\rm spec}^q \:
  I_\Psi
  \nonumber
  \\
  & = &
  \frac{(4.85\times 10^{-4})^s}{2.07\times10^{-9}}
  \frac{(q-1)}{q}
  \sqrt{\frac{(\gamma-1)}{ q r^\gamma } 
        \frac{ q r^\gamma - 1 }{ ( r - 1 ) }} \,
  \frac{I_\Psi}{I_{\rm spec}} \,
  ,
  \nonumber
\end{eqnarray}
where the normalisation constant $C_\Psi$ has been determined by the
condition $\Psi({\cal M} \to \infty ) = 1$. We note that there is also
a weak dependency from the downstream plasma temperature via the
integral $I_{\rm spec}$. However, for weak shocks, ${\cal M} \lesssim
3$, the expression $\Psi$ decreases rapidly, see \Fig{fig-psi}, while
it is virtually constant with $\Psi \sim 1$ for strong shocks. This
implies that rather strong shocks with Mach numbers ${\cal M} \gtrsim 3$ are
necessary to produce any observable radio emission.

After subsuming the terms governed by the strength of the shock under $\Psi(\cal M)$, we rewrite the total emission \Eq{eq-C-all} 
\begin{eqnarray}  
  \frac{{\rm d} P ( \nu_{\rm obs} )}{{\rm d} \nu}
  & = &
  A \:
  n_{\rm e} \:
  C_{\rm spec}^p \:
  C_{\rm sync} \:
  \left( \frac{B}{ \rm \mu G} \right)^\frac{s}{2}
  \left( \frac{\rm 1.4 \: GHz }{ \nu_{\rm obs}} \right)^\frac{s}{2}
  \frac{\sqrt{u\down}}{C_{\rm cool}}
  \frac{1}{C_\Psi} \;
  \Psi(\cal M)
  \nonumber
  \\
  & = &
  6.4 \times 10^{34} \frac{\rm erg}{\rm s \, Hz } \;
  \frac{A}{{\rm Mpc^2}} \,
  \frac{ n_{\rm e}}{\rm 10^{-4} cm^{-3}} \,
  \frac{ \xi_{\rm e} }{0.05 } \:
  \left(\frac{\nu_{\rm obs}}{\rm 1.4 \, GHz} \right)^{-\frac{s}{2}}
  \label{eq-total}
  \\
  && \qquad\qquad
  \times
  \left( \frac{T\down}{\rm 7\, keV} \right)^{\frac{3}{2}} \:
  \frac{ \left( \frac{B}{\rm  \mu G} \right)^{1+\frac{s}{2}} }
        {     \left(\frac{B_{\rm CMB}}{\rm \mu G}\right)^2
            + \left(\frac{B}{\rm \mu G}\right)^2 }
  \;
  \Psi(\cal M)
  .
  \nonumber
\end{eqnarray}
Note that the emission follows a power-law $\propto \nu^{-s/2}$ and thus, for $s=2$ it goes as $\propto \nu^{-1}$.
Since $\Psi({\cal M}) \leq 1 $ and $B^{1+s/2}/(B_{\rm CMB}^2 +B^2)\leq 1$ for $s = 2$, the factor in front,  $6.4 \times 10^{34} \: {\rm erg \, s^{-1} \, Hz^{-1} \, Mpc^{-2}}$, poses an upper limit for the synchrotron emission with respect to the Mach number of the shock and the strength of the magnetic field.

\begin{figure}
  \begin{center}
  \includegraphics[width=0.7\textwidth,angle=-90]{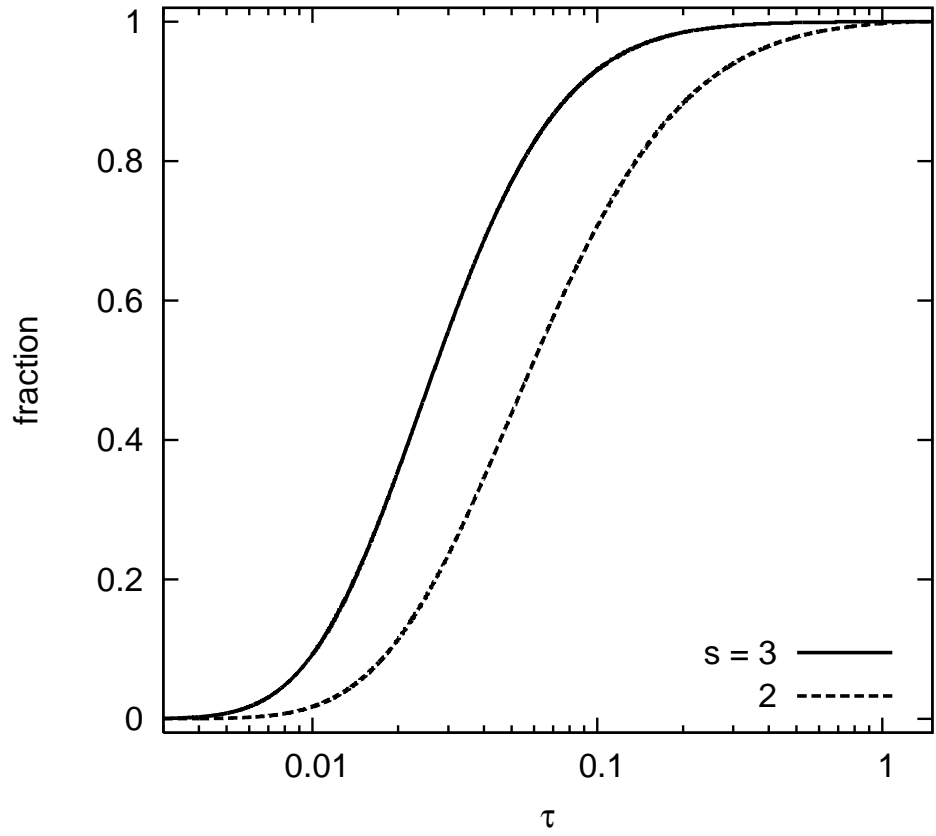}
  \end{center}
  \caption
  { 
  Cumulative synchrotron emission. Computed according to \Eq{eq-cum-synchro} assuming a power-law for the 
  electron energy spectrum, $n_E(\tau) \propto \tau^{-s}$.
  }
  \label{fig-cum-synchro}
\end{figure}

If a radio relic is seen edge-on, its extension perpendicular to the
shock front can be determined. From our model, we can compute at which
distance, $y_{\rm c}$, the synchrotron emission in the downstream
region decreases to a fraction (which we somewhat arbitrarily set to
30\%) of its value immediately behind the shock front. We use such a
rather high fraction since up to 10\,\% of the emission comes from
electrons with $\tau\lesssim 0.01$, see \Fig{fig-cum-synchro}. By the help 
of \Eq{eq-def-tau} we can derive the corresponding Lorentz factor. For instance, 
in a region with a magnetic field of $4\:{\rm \mu G}$ and with an observing frequency
of 1.4\,GHz, 10\,\% of the synchrotron emission is originated by electrons 
with $\gamma_{\rm L} \lesssim 100$.
The spectral electron density in this energy range, however, would be significantly diminished by Coulomb cooling, which is not included in our analysis. Thus, the distance where the emission diminishes to 30\,\% of the initial value seems to be a realistic estimate for the extension of the observable region.

\begin{figure}
  \begin{center}
  \includegraphics[width=0.7\textwidth,angle=-90]{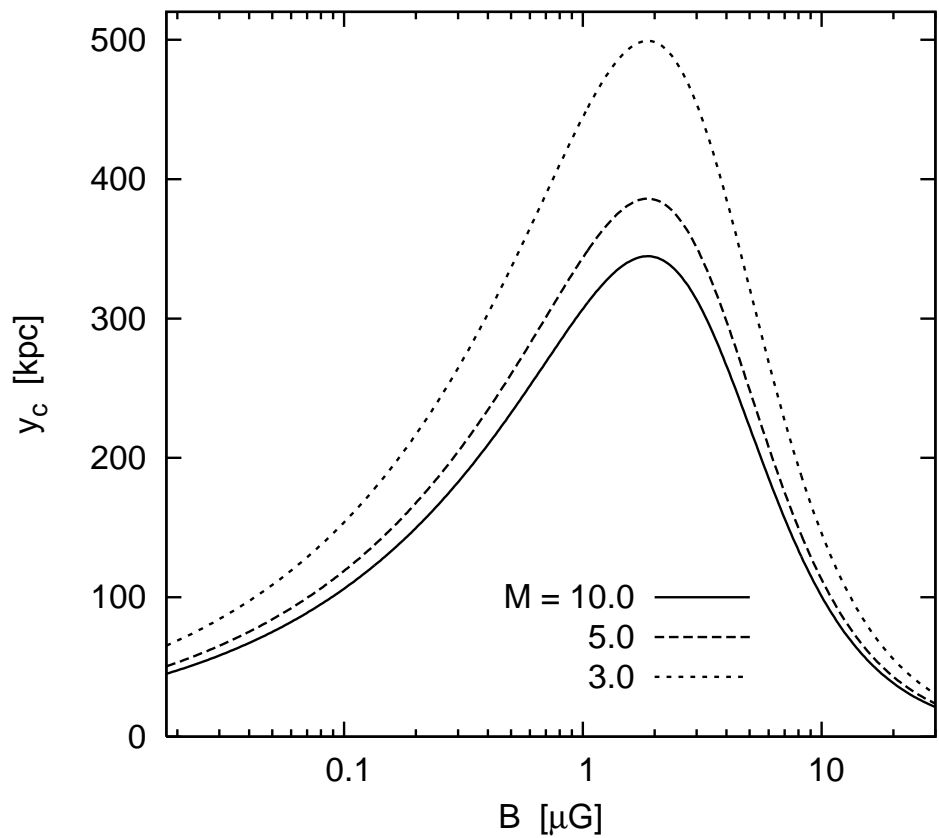}
  \end{center}
  \caption
  {
The distance at which the radio emission decreases to 30\,\% of the value at the shock front. 
  }
  \label{fig-B-dcease}
\end{figure}

To obtain the width of the emission region, we derive first the dimensionless distance $\eta_{\rm c}$, where $\eta$ has been introduced in \Eq{eq-def-eta}. The emission depends on the distance to the shock front, i.\,e. $\eta$, via the integral $\int ... \,{\rm d}\tau$ in \Eq{eq-C-all}. We compute at which $\eta_{\rm c}$ this integral drops to 30\,\% of its value for $\eta=0$. Since the integration depends on the slope of the electron spectrum, $s$, also $\eta_{\rm c}$ does. We find a linear dependence, namely $\eta_{\rm c}(s) = 4.8 \times s - 1.5$. Using the definition of $\eta$ we derive the effective width of the emission region  
\begin{eqnarray} 
  y_{\rm c}
  =
  \frac{C_\tau \, v\down}{C_{\rm cool}} \,
  \eta_{\rm c} 
  =
  120 \: {\rm kpc } \;\;
  \frac{ \left( \frac{B}{\rm  \mu G} \right)^{\frac{1}{2}} }
        { \left( \frac{B_{\rm CMB}}{\rm  \mu G} \right)^2 + \left(\frac{B}{\rm \mu G}\right)^2 } \:
  \left(\frac{\nu_{\rm obs}}{\rm 1.4 \, GHz} \right)^{\frac{s}{2}} \:
  \frac{v\down}{\rm 100 \: km \, s^{-1}}  \:
  \eta_{\rm c} 
  \label{eq-def-yc}
\end{eqnarray}
This equation shows the dependence of the width of the relic, $y_{\rm
c}$, on the magnetic field. In the regime of small magnetic field strengths, $B \ll B_{\rm CMB} = 3.2 \: {\rm\mu G}$ (for $z=0$), $y_{\rm
c}$ is proportional to $B^{1/2}$. Higher magnetic fields produce
thicker relics because a higher magnetic field lowers the relevant
Lorentz factor for synchrotron emission. In contrast, in a regime with
strong magnetic field, $B\gg  B_{\rm CMB}$, the $y_{\rm c}$ shrinks
with increasing B-field according to $y_{\rm c} \propto B^{-3/2}$ due to the synchrotron cooling itself. The width, $y_{\rm c}$, is maximal for $B \sim  2 \:{\rm \mu G}$, see \Fig{fig-B-dcease}.

Also the strength of the shock affects the size of the synchrotron
emitting area. Rather weak shocks, ${\cal M} \lesssim 3$, are more
extended, since the spectrum is steeper. For a spectral index of
$s=3$, electrons with smaller $\tau$ are mainly responsible for the
emission than for $s=2$, see \Fig{fig-cum-synchro}. Hence it takes longer to diminish
the density of the relevant electron population. For small
$\tau$, i.\,e., weak shocks and strong magnetic fields, the electron
population is strongly reduced by Coulomb cooling, which is not
included in our model. For those shocks the extension of the emission
area is smaller than predicted here.

We can also derive the time after which the emission of the downstream plasma has dropped to 30\,\% of its initial value from \Eq{eq-def-yc},
\begin{eqnarray} 
  t_{\rm c}
  =
  \frac{y_{\rm c}}{v\down}
  =
  \frac{C_\tau \, v\down}{C_{\rm cool}}
  .
  \nonumber
\end{eqnarray}
It has a maximum at $B \sim  2 \:{\rm \mu G}$, which amounts to $\sim 1 \: {\rm Gyr}$ for ${\cal M} = 4$.

The code for computing the synchrotron emission and the width of the emission region will be provided upon request.

\section{Cluster radio relic examples}

\label{sec-examples}

\subsection{Abell-3667}

\label{sec-A3667}

A textbook example of a merging cluster with strong shocks traveling
  into the cluster periphery is Abell\,3667. This cluster is unique
  since it shows prominent radio emission \citep{roettgering:97} which
  is presumably caused by an outgoing shock wave. In X-rays, there is
  a double peak in the cluster centre \citep{shibata:99} and a cold
  front indicating a fast moving substructure
  \citep{markevitch:02}. Both, indicates that the cluster suffered a
  merger recently. Even if the X-ray flux from the cluster periphery
  is not sufficient to identify a shock front at the position of
  relics, \citet{roettiger:99} showed that X-ray morphology and
  position of the relics is consistent with a merger of a 20\,\%
  substructure 1\,Gyr ago. The spectral index of the north-west relic
  varies clearly: It steepens from $\alpha \sim 0.6$ at the outer rim
  to 1.5 at its faint rim towards the cluster centre. This suggests
  that the relic is basically seen edge-on: The outer rim is populated by
  recently accelerated electrons, while towards the centre, the
  electrons population has aged.

A similar steepening of the electron spectrum is expected in our
 model. As the plasma moves
 downstream and the electrons age, the radio spectrum steepens. However, the overall
 spectrum is a power-law, see \Eq{eq-total}, even if locally the
 spectrum may steepen towards higher frequencies. Note that the
 power-law is only obtained if all contributions to the total emission
 are present: from the newly accelerated electron population at the
 shock front up to the old electrons that are too cold to emit
 significant synchrotron radiation. In order to apply our model to
 this relic, the downstream region should be homogeneous, i.\,e. it
 should not show adiabatic contraction in this region, etc.. The
 north-west relic in Abell\,3667 is quite close to such an ideal
 scenario. \citet{roettgering:97} found an overall index of $\alpha =
 1.1$. They argued that there is no spectral steeping, what may
 indicate that we see plasma from the injection to the point where the
 emission has faded. Using \Eq{eq-total}, we conclude that the initial slope in the
 electron spectrum is $s=2.2$, which is consistent with $\alpha \sim
 0.6$, close to the shock surface (For a homogenous distribution of
 suprathermal electron with power-law spectral slope, $s$, the radio
 emission has the spectral index $\alpha = (s-1)/2$). By combining \Eq{eq-r-from-q},
 (\ref{eq-mach-from-q-r}) and (\ref{eq-s-from-r}), we can also
 infer the Mach number of the shock. We find that the
 Mach number of the shock related to the north-west relic in A\,3667
 is ${\cal M} = 4.7$.

\begin{figure}
  \begin{center}
  \includegraphics[width=0.7\textwidth,angle=-90]{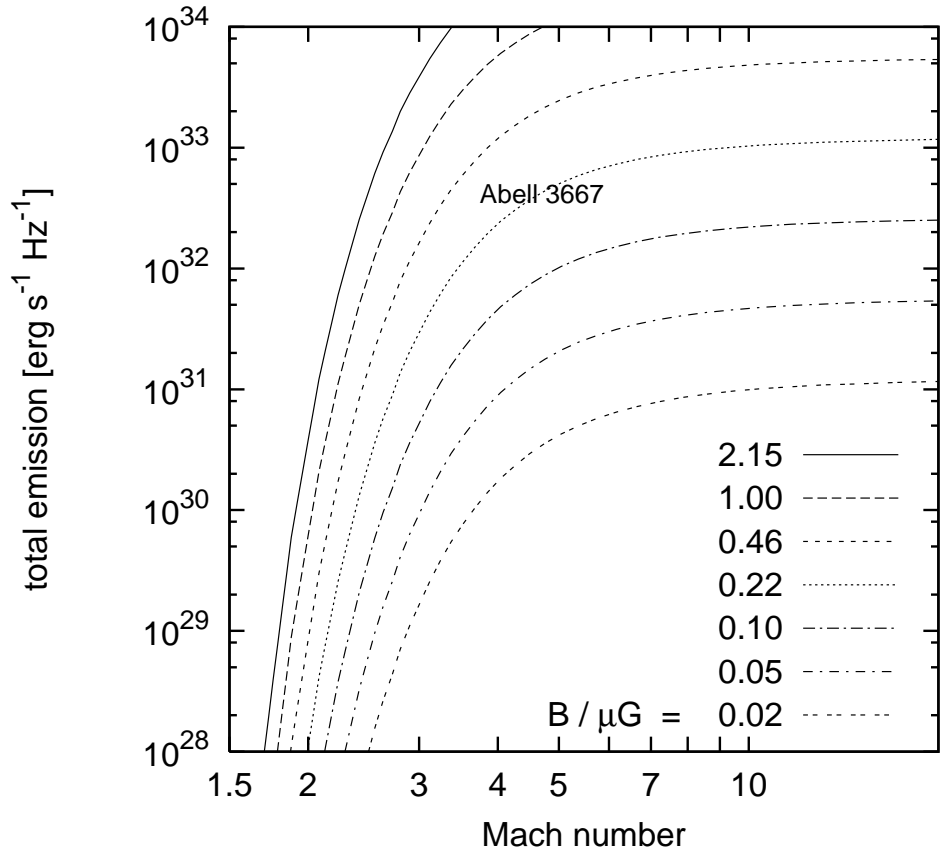}
  \end{center}
  \caption
  { 
Radio power as a function of Mach number for different magnetic field strengths. We have marked the position of the radio power and the derived Mach number for Abell\,3667.
  }
  \label{fig-abell-B}
\end{figure}

We can use our expression for the total emission to estimate the field
 strength in the downstream region. \citet{roettgering:97} derived the
 temperature and electron density in the relic region, $10^8\:{\rm K}$
 and $10^{-4}\:{\rm cm^{-3}}$, respectively. \citet{sarazin:99} argued
 that about 1\,\% of the thermal energy in the ICM is found in
 relativistic electrons, with $\gamma_{\rm L} > 300$. Since our
 definition of $\xi_{\rm e}$ includes also supra-thermal electron with
 $\gamma_{\rm L} < 300$, we choose a somewhat higher value of
 $\xi_{\rm e} = 0.05$. In order to compute the total emission, we also
 need the area of the shock surface. From \citet{roettgering:97},
 Fig.\,4, we estimate that the bulk of radio emission is extended over
 $\sim 2.0\:{\rm Mpc}$. Using these ICM parameters, we derive the
 radio emission as function of the shock strength and the downstream
 magnetic field, see \Fig{fig-abell-B}. The observed total emission is
 $P_{1.4} = 4.1 \times 10^{32} \: {\rm erg \, s^{-1} \, Hz^ {-1}}$.
 We have derived the Mach number of the shock by the spectral index of
 the radio emission. By comparing the observed radio power to the
 computed emission, see \Fig{fig-abell-B}, we conclude that the
 strength of the magnetic field in the relic region is $B\sim
 0.2\:{\rm \mu G}$.

Since this relic is basically seen edge-on, its extension
perpendicular to the shock front, i.\,e. radial with respect to the
cluster centre, can be used to infer the magnetic field strength. In
\Sec{sec-total}, we have introduced a width $y_c$, which is the
shortest distance between the shock front and the downstream position
where the radio emission has decreased to 30\% of its value at the
shock front. The temperature of the downstream plasma and the Mach
number of the shock determine the dimensionless quantity $\eta$ and
the downstream velocity, $v\down$. For the parameters of A3667, we
obtain 9.0 and $720\:{\rm km \, s^{-1}}$, respectively. With a
magnetic field of $0.2\:{\rm \mu G}$ we derive by \Eq{eq-def-yc} that
the width $y_c$ amounts 0.3\,Mpc, which is consistent with the
observed extension of the relic.

  If the gas in the downstream region is substantially compressed, the
  relativistic electrons are re-energized by adiabatic compression.  For a
  polytropic gas with $\gamma = 5/3$, the gas density
  in the periphery of cluster scales with $\rho_{\rm gas} \propto ( \ln(1 +r/r_s) / (r/r_s) )^{3/2}$,
  where $r$ is the distance from the cluster centre and  
  $r_s$ is the scaling radius of the cluster \citep{ascasibar:06}. 
  Most relics are situated on the periphery of galaxy clusters. The radio
  relics in A3667, for example, are located at distances of $\sim 2.6$ Mpc on
  either side from the cluster centre.
  Given that the relics have a width of around $300$ kpc and assuming that the
  downstream gas density follows the density profile of the cluster, the density
  increase along the relic is no more than $\sim 15\,\%$. Such a small
  compression has a moderate effect on the average electron energy, c.f. \citet{ensslin:01}. 
  Flux conservation leads to $B\propto \rho^{2/3}$, hence the magnetic field can be enhanced maximally 
  a few ten percent. Hence, we neglect the effects of adiabatic compression in
  the following discussion.

\subsection{Abell-115}

\label{sec-A115}

Another example of a very extended radio relic has been found in
Abell~155. Its size tangential to the cluster X-ray emission is more
than 2\,Mpc. In contrast, it is very thin in radial direction, a few
hundred kpc at most. Compared to A3667, it has a much lower
radial/tangential extension ratio. Obviously, such a thin radio
structure can only be achieved when the line-of-sight is tangential to
the shock surface. Another prerequisite is that the radio emission
fades quickly while the plasma moves downstream. Considering the
discussion above, this implies a low magnetic field strength in the
downstream plasma.

Similar to the procedure used for the north-west relic in A3667, we can
  infer the magnetic field strength in the emission region. We adopt
  the values given by \citet{shibata:99} for A115. The cluster
  temperature in the periphery is about 5\,keV, and the electron
  density of the order of $10^{-4} \:{\rm cm^{-3}}$. \citet{govoni:01}
  investigated the radio emission and found a total emission of
  $P_{1.4} = 1.9 \times 10^{31} \: {\rm erg \, s^{-1} \, Hz^{-1}}$
  with a slope of $\alpha = 1.1$. We use those values and keep the
  other parameters to the values given above, e.g. the energy fraction
  in suprathermal electrons $\xi_{\rm e}$. Since the spectral slope is
  the same as in A3667 the Mach number of the shock is again ${\cal M}
  = 4.7$. The lower temperature in the downstream region also lowers
  the radio emission. However, the large difference to A3667 in radio
  power can only be achieved if the magnetic field is lower.  We find
  that $B \sim 0.1 \:{ \rm \mu G}$ reproduces the observed emission
  well.

In the direction perpendicular to the shock front, the emission has
dropped to 30\,\% of its initial value at a distance of 0.1\,Mpc from
the shock front. This value is also consistent with the observations.

\subsection{Abell-2256}

\label{sec-A2256}

Abell\,2256 shows a radio halo and relic \citep{clarke:06}. The relic
is seen close to face-on, hence the spectral steepening is only
mild. The average slope is $\alpha = - 1.2$, which indicates a shock
with Mach number ${\cal M} = 3.3$. From \citet{rephaeli:03} we adopt a
cluster temperature of 7.8~keV.  The observed extension of the relic is about
1~Mpc, hence we assume a shock surface area of $1\,Mpc^2$. The power,  $P_{1.4} =
3.6 \times 10^{31} \: {\rm erg \, s^{-1} \, Hz^{-1}}$ is reproduced
provided the magnetic field has a strength of $B
\sim 0.3 \:{ \rm \mu G}$.

This value for the magnetic field is almost a magnitude lower as that
in the central halo region (see \citet{clarke:06}) and indicates that
the magnetic field decreases with increasing radius from the centre.

If the radio relic is being produced by a giant shock wave, what has
produced the radio halo in Abell\,2256? Can the two phenomena be
linked?
With the magnetic field in the plasma also the relevant Lorentz factor
for the radio emission changes. If the magnetic field in the halo
region is about a few $\mu$G as indicated by \citet{clarke:06}, the
Lorentz factor of those electrons that produce the radio emission is
$\gamma_{\rm L, half} \sim 300$. In contrast, in the relic region the
lower magnetic field implies that the Lorentz factor of the
radio-emitting electrons is $\gamma_{\rm L, half} \sim 1000$. At the
field strengths in both relic and halo, we can assume that the cooling
of the electron spectrum is mainly causes by inverse-Compton
cooling. Since the cooling time is inversely proportional to
$\gamma_{\rm L}$, the cluster core would emit synchrotron radiation
three times longer. Hence it is conceivable that the halo and relic
are caused by the same shock front.

\section{Radio emission versus cluster temperature} 

\label{sec-P-T}

The example clusters discussed above indicate that shocks which 
produce presently observable radio emission have Mach numbers of about ${\cal M} \sim
3-5$. It is interesting to note that significantly stronger shocks
show only moderately more radio emission if all other downstream
parameters are the same, as is shown in \Fig{fig-psi}. This is due to the fact
that strong shocks reach an asymptotic slope for the electron
spectrum. Much weaker shocks would have virtually no radio
emission. \citet{gabici:03} found that it is quite unlikely to achieve
shock strength of ${\cal M} \gtrsim 4$ in a cluster merger. The
combination of the paucity of shocks with ${\cal M} \gtrsim 4$ and the
strong suppression of radio emission in weaker shocks may explain to
some extent why only so few radio relics have been found.

\begin{figure}
  \begin{center}
  \includegraphics[width=0.7\textwidth,angle=-90]{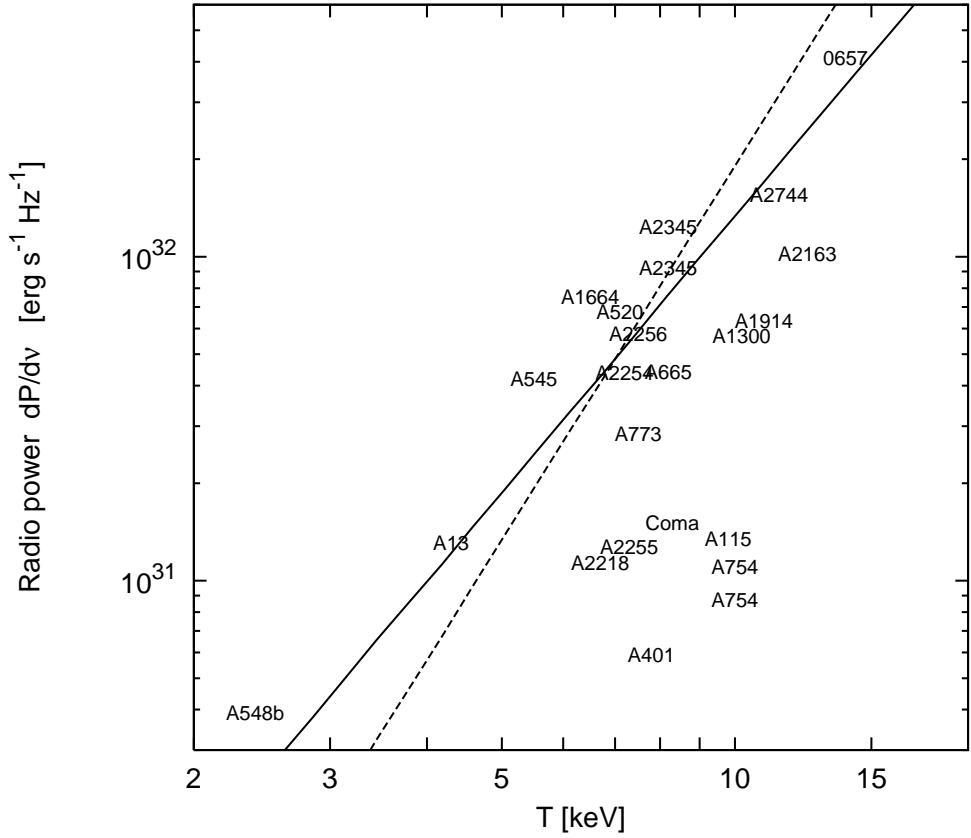}
  \end{center}
  \caption
  {
Radio power for an observing frequency of 1.4\,GHz versus cluster temperature. The solid line indicates the radio power for a shock front with an area of 1\,Mpc$^2$, while the dashed line is computed for a temperature dependent area, $A\propto T$, see \Fig{fig-T-LLS}. Radio and X-ray data for the clusters are taken from \citet{giovannini:99}, \citet{govoni:01}, and  \citet{govoni:04b}.
  }
  \label{fig-I-temp}
\end{figure}

If we consider the magnetic field dependency in \Eq{eq-total}, one
 striking feature is that for strong field strengths, $B \gg B_{\rm
 CMB}$, the emission is constant. Hence, even an arbitrarily strong
 shock with an arbitrarily strong magnetic field would not exceed an
 emission of $\sim 6 \times 10^{34} \:{\rm erg \, s^{-1} \, Hz^{-1}}$,
 with standard parameters for temperature and density. Thus, there is
 a distinct upper limit for the radio emission. Intriguingly, our few
 examples support the assumption that the magnetic field in the ICM is
 correlated with its temperature. We found that A3667 and A2256 which
 have temperatures of about 8\,keV lead to magnetic field strengths of 0.2-0.3\,$\mu$G. In contrast, A115 with a temperature of 5\,keV shows a field strength of 0.1\,$\mu$G in the relic region. However, we restrict ourselves here to the custom assumption that a fraction, $\xi_B$, of the energy dissipated at the shock front is converted into magnetic fields,
\begin{eqnarray}  
  \frac{B^2}{8\pi}
  =
  \xi_B \,
  m_{\rm p}
  n_e
  u
  .
  \label{eq-B-equi}
\end{eqnarray}
On average, the magnetic field energy fraction is $\langle \xi_B
\rangle = 6\times10^{-4}$ for our example clusters, see
Tab.\,\ref{tab-abell}. Thus, only a very small fraction of the
dissipated energy has to be converted. For A115 and A2256, the values
of $\xi_B$ may be underestimated to some extent, since we have assumed
that, at the location of the relics, the plasma has the average cluster temperature. Only for A3667, a temperature estimate for the relic region is available. However, we use \Eq{eq-B-equi} to compute the radio power as a function of the cluster temperature, see \Fig{fig-I-temp}, solid line. Here we have assumed that the typical strength of a strong shock in the ICM is ${\cal M} = 4$ and that the length scale of the shock fronts is 1\,Mpc.

\begin{figure}
  \begin{center}
  \includegraphics[width=0.7\textwidth,angle=-90]{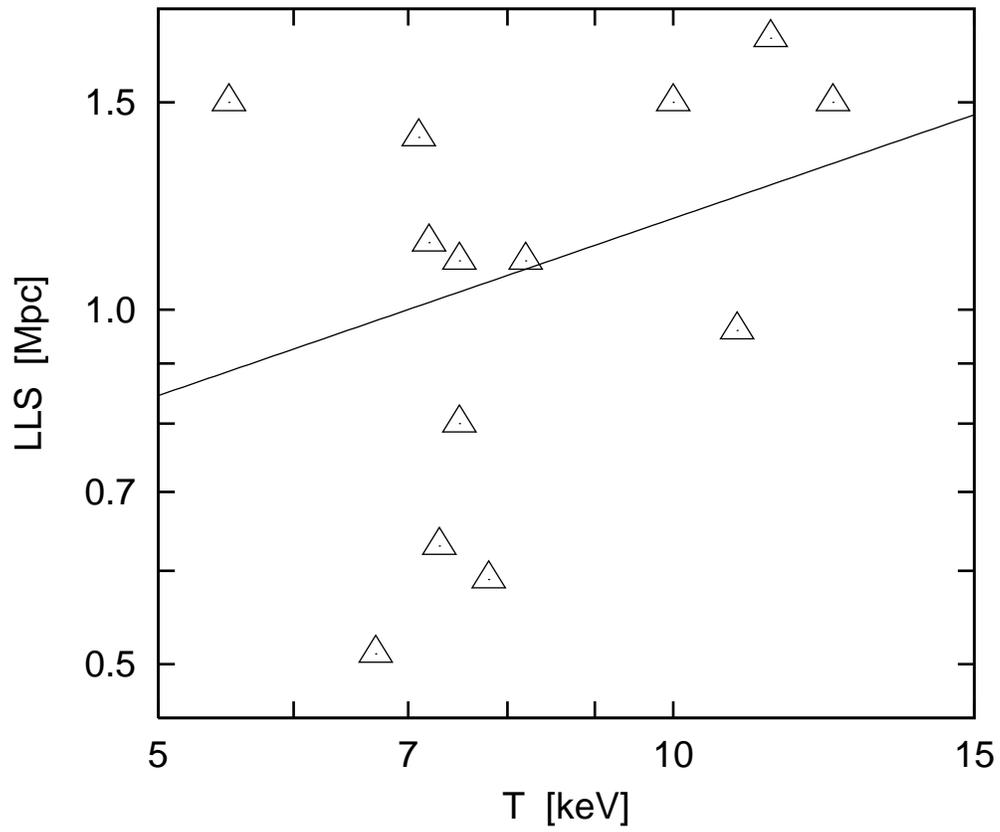}
  \end{center}
  \caption
  { 
Longest linear size versus cluster temperature for radio halos shown in \Fig{fig-I-temp}. The solid line indicates ${\rm LLS} = 1\:{\rm Mpc} ( T / 7\:{\rm keV})^{1/2} \: T$.
  }
  \label{fig-T-LLS}
\end{figure}

Our estimate for the radio emission is close to the emission observed for radio halos. This may indicate that a significant part of the radio halo emission may be the result of an earlier merger shock. The observed radio power-temperature relation seems to be steeper than our result. This may be attributed to scaling relations that we have ignored in our analysis. It seems plausible that the linear size of radio halos depends on the cluster temperature, see \Fig{fig-T-LLS}. In average one may expect that the extension of halos is a fraction of the virial radius, $r_{\rm vir}$ which scales with the cluster temperature by $r_{\rm vir} \propto M^{1/3} \propto T^{1/2}$ \citep[for cluster scaling relations see e.\,g.][]{eke:98}. Including the linear size-temperature relation into our estimate for the radio power, we obtain a somewhat steeper dependency, see \Fig{fig-I-temp}. The slope of the radio power is now ${\rm d}P/{\rm d}\nu \propto T^{(12+s)/4}$, see \Eq{eq-total}, where we
  
 have used $B^2 \propto T$ and $A\propto T$. Hence, the strong temperature dependence is caused by the combination of several facts: With increasing cluster temperature, the electron density in the relevant Lorentz factor regime increases, the magnetic field scales with the temperature, and the average linear size of the emission increases with cluster temperature, i.\,e. its mass.

\section{Cosmic accretion shocks}

\label{sec-cosmic}

Numerical simulations suggest that structure formation shocks may be subdivided into two classes \citep{ryu:03}. Merger (or `internal') shocks are caused by the supersonic motion of infalling substructures, as seen, e.\,g., in A\,3667 or in the `bullet cluster' 1E\,0657-56 \citep{markevitch:02}. Typically, the Mach number of these shocks is rather low, ${\cal M} \lesssim 3$, \citep{gabici:03}, since the cluster temperature reflects already the depth of the gravitational potential which also determines the maximal velocities in the course of the merger. When the shocks travel outwards into the cluster surroundings, the Mach number increases since the gas there has virtually never been shock-heated and is therefore much colder. Numerical simulations indicate that those accretion (or `external') shocks reach Mach numbers $\gg 10$ \citep{miniati:00, ryu:03, pfrommer:06, hoeft:06}. Basically, clusters are enveloped by a sphere of accretion shocks at a distance, $r_{\rm sh}$, of a 
 few times the virial radius, $r_{\rm vir}$, \citep{nagai:03, ryu:03, hoeft:06}. Due to the accretion shock, the temperature of the gas rises from $\sim 10^3 - 10^4 \:{\rm K}$ to almost the cluster temperature. To obtain a rough estimate for the temperature in the cluster periphery, we utilize the profile given by \citet{loken:02}
\begin{eqnarray}
  T
  =
  1.3 \:
  \langle T \rangle \:
  \left(
    1
    +
    1.5 \,
    \frac{r}{r_{\rm vir}}
  \right)^{-1.6}
  ,
  \nonumber
\end{eqnarray}
where $\langle T \rangle$ is the emission-weighted temperature of the cluster. With $r_{\rm sh} \sim 3 r_{\rm vir}$ the temperature in the downstream region of accretion shocks is $\sim 0.09 \, \langle T \rangle$. Hence, we can determine the total emission caused by accretion shocks surrounding clusters of galaxies. For a cluster with $\langle T \rangle = 20\,{\rm keV}$ and $ r_{\rm vir} = 2\:{\rm Mpc}$, we estimate the temperature in the downstream region of the accretion shocks, namely $1.7\,{\rm keV}$. Assuming that the electron density at such a large distance from the cluster centre amounts to $10^{-6}\:{\rm cm^{-3}}$, we derive the magnetic field by \Eq{eq-B-equi}, we obtain $0.01\,{\rm\mu G}$. With the help of \Eq{eq-total}, we compute the total emission of a sphere with $12\:{\rm Mpc}$ diameter. We obtain for a strong shock, ${\cal M} \gg 10$,
\begin{eqnarray}
  \left.
  \frac{{\rm d} P ( 1.4\:{\rm GHz} )}{{\rm d} \nu}      
  \right|_{\rm acc-sh}
  \sim 
  2.2 \times 10^{29}\:{\rm erg \, s^{-1} \, Hz^{-1}}
  .
  \nonumber
\end{eqnarray}
In reality, the accretion shock geometry is much more complex and projection effects have to be considered. Therefore, a significant fraction of the emission may come from a small region. Note that for such weak magnetic fields the emission region has only a small extension perpendicular to the shock front ($\sim 50\:{\rm kpc}$).

Upcoming radio telescopes such as {\sc Lofar} and SKA will allow to
explore the radio sky with significantly improved sensitivity and
resolution in the frequency range from 10\,MHz to 1\,GHz.
\citet{keshet:04} inferred that the radio emission from structure
formation shocks may become detectable with these telescopes. Hence,
in near future we will have much stronger constraints for modeling the
relativistic electron populations in clusters of galaxies.

If we assume that 3\,\% of the emission comes from a rather compact
region, we expect at 30\,MHz a spot-like emission of $\sim
4\times10^{29}\:{\rm erg \, s^{-1} \, Hz^{-1}}$ (note
$P\propto\nu^{-s/2}$). If this cluster is located at $z=0.1$ the flux
would amount to $1\:{\rm mJy}$. This is close to the sensitivity of
{\sc Lofar}, hence accretion shocks with optimal conditions for radio
emission will presumably be observed with sensitive radio telescopes
in near future.

\section{Summary}

\label{sec-summary}

Galaxy clusters grow either by mergers or by steady accretion, both of
 which produce large shock fronts. In particular, the extended radio
 relics ({\it radio gischt}) in the periphery of clusters are believed
 to trace these fronts.  In analogy to supernova remnants, diffuse
 shock acceleration may produce a population of relativistic electrons
 that cool in the downstream region of the shock by inverse Compton
 and synchrotron losses. Here, we compute the total emission in the
 downstream region per unit area of the shock surface. The resulting
 analytic expression shows several interesting features: (i) the total
 spectrum is a power-law even if locally the spectrum may steepen
 towards higher frequencies. The overall slope is close to unity and
 depends solely on the slope of the initial electron spectrum which,
 in turn, is directly related to the strength of the shock. Thus, the
 overall slope of the radio emission may allow us to infer the Mach
 number of the shock. (ii) the radio emission depends almost like a
 step-function on the shock strength. For Mach numbers $\lesssim 3$,
 only very little radiation is emitted while for ${\cal M} \gtrsim 10$
 the emission saturates.  (iii) in the regime of weak magnetic fields,
 $B\ll B_{\rm CMB}$, the total emission is proportional to $B^{1+s/2}$
 while for strong magnetic fields the total emission is virtually
 independent of $B$. (iv) The width of the emission region,
 i.\,e. the extension perpendicular to the shock front, has a maximum
 extension for $B\sim B_{\rm CMB}$.

We analysed the relics in A\,115, A\,2256, and A\,3667 and found that
the shocks have Mach numbers in the range of 3 to 5. Weaker shocks
would lead to practically no radio emission. Furthermore, we can infer
the magnetic fields in the emission regions. We find that for three
sample relics the magnetic field is in the range from 0.1 to
$0.3\:{\rm\mu G}$. The extension of the relics perpendicular to the
shock front serves as an independent indicator for the field
strength. The energy fraction stored in the magnetic field seems to be
rather low, namely $\xi_B \sim 10^{-3}$.

Assuming that the magnetic field energy density is on average
proportional to the thermal one, we derive a radio
emission-temperature relation. It agrees strikingly well with that
found for radio halos. Moreover, if in the centres
of galaxy clusters, i.\,e. where the radio halos reside, the magnetic
fields is about a
few ${\rm\mu G}$, the extension of the emission region perpendicular
to the shock front would be several hundred kpc. Therefore, it is
conceivable that at least part of the halo emission stems from shock
waves which swept over the cluster centres within the last 1\,Gyr. Using the
value obtained for $\xi_B$, we have finally estimated the emission
from cosmic accretion shocks. We inferred that at 30\,MHz a radio
power of $10^{29}$ - $10^{30}\:{\rm erg \, s^{-1} \,Hz^{-1}}$ may be
expected from regions where the accretion shock surface is
parallel to the line-of-sight. A detailed computation of the cosmic
radio background will be subject of a forthcoming study.
\\

\noindent {\sc Acknowledgement }
We gratefully acknowledge support by DFG grant BR 2026/2. We thank
Torsten Ensslin for helpful discussions.

\newcommand{\aap  }{A\&A}
\newcommand{\araa }{ARA\&A}
\newcommand{\apj  }{ApJ}
\newcommand{\apjs }{ApJS}
\newcommand{\apjl }{ApJL}
\newcommand{\apss }{ApSS}
\newcommand{\aapr }{A\&A~Rev.}
\newcommand{\aj   }{AJ}
\newcommand{\jgr  }{JGR}
\newcommand{\mnras}{MNRAS}
\newcommand{\physrep  }{PhysRep}

\bibliography{radio}

\begin{thebibliography}{}

\bibitem[\protect\citeauthoryear{{Amato} \& {Blasi}}{{Amato} \&
  {Blasi}}{2005}]{amato:05}
{Amato} E.,  {Blasi} P.,  2005, \mnras, 364, L76

\bibitem[\protect\citeauthoryear{{Ascasibar}, {Sevilla}, {Yepes}, {M{\"u}ller}
  \& {Gottl{\"o}ber}}{{Ascasibar} et~al.}{2006}]{ascasibar:06}
{Ascasibar} Y.,  {Sevilla} R.,  {Yepes} G.,  {M{\"u}ller} V.,
  {Gottl{\"o}ber} S.,  2006, \mnras, 371, 193

\bibitem[\protect\citeauthoryear{{Axford}, {Leer} \& {Skadron}}{{Axford}
  et~al.}{1978}]{axford:78}
{Axford} W.~I.,  {Leer} E.,    {Skadron} G.,  1978, in International Cosmic Ray
  Conference {The acceleration of cosmic rays by shock waves}.
pp 132--137

\bibitem[\protect\citeauthoryear{{Bell}}{{Bell}}{1978a}]{bell:78a}
{Bell} A.~R.,  1978a, \mnras, 182, 147

\bibitem[\protect\citeauthoryear{{Bell}}{{Bell}}{1978b}]{bell:78b}
{Bell} A.~R.,  1978b, \mnras, 182, 443

\bibitem[\protect\citeauthoryear{{Bennett} \& {Ellison}}{{Bennett} \&
  {Ellison}}{1995}]{bennett:95}
{Bennett} L.,  {Ellison} D.~C.,  1995, \jgr, 100, 3439

\bibitem[\protect\citeauthoryear{{Berezhko}, {Ksenofontov} \&
  {V{\"o}lk}}{{Berezhko} et~al.}{2003}]{berezhko:03}
{Berezhko} E.~G.,  {Ksenofontov} L.~T.,    {V{\"o}lk} H.~J.,  2003, \aap, 412,
  L11

\bibitem[\protect\citeauthoryear{{Blandford} \& {Eichler}}{{Blandford} \&
  {Eichler}}{1987}]{blandford:87}
{Blandford} R.,  {Eichler} D.,  1987, \physrep, 154, 1

\bibitem[\protect\citeauthoryear{{Blandford} \& {Ostriker}}{{Blandford} \&
  {Ostriker}}{1978}]{blandford:78}
{Blandford} R.~D.,  {Ostriker} J.~P.,  1978, \apjl, 221, L29

\bibitem[\protect\citeauthoryear{{Brunetti}}{{Brunetti}}{2004}]{brunetti:04}
{Brunetti} G.,  2004, Journal of Korean Astronomical Society, 37, 493

\bibitem[\protect\citeauthoryear{{Clarke} \& {Ensslin}}{{Clarke} \&
  {Ensslin}}{2006}]{clarke:06}
{Clarke} T.~E.,  {Ensslin} T.,  2006, ArXiv Astrophysics e-prints

\bibitem[\protect\citeauthoryear{{Colafrancesco} \& {Blasi}}{{Colafrancesco} \&
  {Blasi}}{1998}]{colafrancesco:98}
{Colafrancesco} S.,  {Blasi} P.,  1998, Astroparticle Physics, 9, 227

\bibitem[\protect\citeauthoryear{{Dahle}, {Kaiser}, {Irgens}, {Lilje} \&
  {Maddox}}{{Dahle} et~al.}{2002}]{dahle:02}
{Dahle} H.,  {Kaiser} N.,  {Irgens} R.~J.,  {Lilje} P.~B.,    {Maddox} S.~J.,
  2002, \apjs, 139, 313

\bibitem[\protect\citeauthoryear{{Dennison}}{{Dennison}}{1980}]{dennison:80}
{Dennison} B.,  1980, \apjl, 239, L93

\bibitem[\protect\citeauthoryear{{Drury}}{{Drury}}{1983}]{drury:83}
{Drury} L.~O.,  1983, Reports of Progress in Physics, 46, 973

\bibitem[\protect\citeauthoryear{{Dyer}, {Reynolds}, {Borkowski}, {Allen} \&
  {Petre}}{{Dyer} et~al.}{2001}]{dyer:01}
{Dyer} K.~K.,  {Reynolds} S.~P.,  {Borkowski} K.~J.,  {Allen} G.~E.,    {Petre}
  R.,  2001, \apj, 551, 439

\bibitem[\protect\citeauthoryear{{Eke}, {Navarro} \& {Frenk}}{{Eke}
  et~al.}{1998}]{eke:98}
{Eke} V.~R.,  {Navarro} J.~F.,    {Frenk} C.~S.,  1998, \apj, 503, 569

\bibitem[\protect\citeauthoryear{{Ellison} \& {Cassam-Chena{\"i}}}{{Ellison} \&
  {Cassam-Chena{\"i}}}{2005}]{ellison:05}
{Ellison} D.~C.,  {Cassam-Chena{\"i}} G.,  2005, \apj, 632, 920

\bibitem[\protect\citeauthoryear{{En{\ss}lin}, {Biermann}, {Klein} \&
  {Kohle}}{{En{\ss}lin} et~al.}{1998}]{ensslin:98}
{En{\ss}lin} T.~A.,  {Biermann} P.~L.,  {Klein} U.,    {Kohle} S.,  1998, \aap,
  332, 395

\bibitem[\protect\citeauthoryear{{En{\ss}lin} \& {Br{\" u}ggen}}{{En{\ss}lin}
  \& {Br{\" u}ggen}}{2002}]{ensslin:02}
{En{\ss}lin} T.~A.,  {Br{\" u}ggen} M.,  2002, \mnras, 331, 1011

\bibitem[\protect\citeauthoryear{{En{\ss}lin} \& {Gopal-Krishna}}{{En{\ss}lin}
  \& {Gopal-Krishna}}{2001}]{ensslin:01}
{En{\ss}lin} T.~A.,  {Gopal-Krishna} 2001, \aap, 366, 26

\bibitem[\protect\citeauthoryear{{Fabian}, {Reynolds}, {Taylor} \&
  {Dunn}}{{Fabian} et~al.}{2005}]{fabian:05}
{Fabian} A.~C.,  {Reynolds} C.~S.,  {Taylor} G.~B.,    {Dunn} R.~J.~H.,  2005,
  \mnras, 363, 891

\bibitem[\protect\citeauthoryear{{Feretti}}{{Feretti}}{2005}]{feretti:05}
{Feretti} L.,  2005, Advances in Space Research, 36, 729

\bibitem[\protect\citeauthoryear{{Feretti}, {Brunetti}, {Giovannini}, {Kassim},
  {Orr{\'u}} \& {Setti}}{{Feretti} et~al.}{2005}]{feretti:05b}
{Feretti} L.,  {Brunetti} G.,  {Giovannini} G.,  {Kassim} N.,  {Orr{\'u}} E.,
   {Setti} G.,  2005, Journal of Korean Astronomical Society, 37, 315

\bibitem[\protect\citeauthoryear{{Feretti}, {Orr{\`u}}, {Brunetti},
  {Giovannini}, {Kassim} \& {Setti}}{{Feretti} et~al.}{2004}]{feretti:04}
{Feretti} L.,  {Orr{\`u}} E.,  {Brunetti} G.,  {Giovannini} G.,  {Kassim} N.,
   {Setti} G.,  2004, \aap, 423, 111

\bibitem[\protect\citeauthoryear{{Fujita} \& {Sarazin}}{{Fujita} \&
  {Sarazin}}{2001}]{fujita:01}
{Fujita} Y.,  {Sarazin} C.~L.,  2001, \apj, 563, 660

\bibitem[\protect\citeauthoryear{{Gabici} \& {Blasi}}{{Gabici} \&
  {Blasi}}{2003}]{gabici:03}
{Gabici} S.,  {Blasi} P.,  2003, \apj, 583, 695

\bibitem[\protect\citeauthoryear{{Giovannini}, {Tordi} \&
  {Feretti}}{{Giovannini} et~al.}{1999}]{giovannini:99}
{Giovannini} G.,  {Tordi} M.,    {Feretti} L.,  1999, New Astronomy, 4, 141

\bibitem[\protect\citeauthoryear{{Govoni}, {En{\ss}lin}, {Feretti} \&
  {Giovannini}}{{Govoni} et~al.}{2001}]{govoni:01b}
{Govoni} F.,  {En{\ss}lin} T.~A.,  {Feretti} L.,    {Giovannini} G.,  2001,
  \aap, 369, 441

\bibitem[\protect\citeauthoryear{{Govoni} \& {Feretti}}{{Govoni} \&
  {Feretti}}{2004}]{govoni:04}
{Govoni} F.,  {Feretti} L.,  2004, International Journal of Modern Physics D,
  13, 1549

\bibitem[\protect\citeauthoryear{{Govoni}, {Feretti}, {Giovannini},
  {B{\"o}hringer}, {Reiprich} \& {Murgia}}{{Govoni} et~al.}{2001}]{govoni:01}
{Govoni} F.,  {Feretti} L.,  {Giovannini} G.,  {B{\"o}hringer} H.,  {Reiprich}
  T.~H.,    {Murgia} M.,  2001, \aap, 376, 803

\bibitem[\protect\citeauthoryear{{Govoni}, {Markevitch}, {Vikhlinin},
  {VanSpeybroeck}, {Feretti} \& {Giovannini}}{{Govoni}
  et~al.}{2004}]{govoni:04b}
{Govoni} F.,  {Markevitch} M.,  {Vikhlinin} A.,  {VanSpeybroeck} L.,  {Feretti}
  L.,    {Giovannini} G.,  2004, \apj, 605, 695

\bibitem[\protect\citeauthoryear{{Hoeft}, {Br{\"u}ggen} \& {Yepes}}{{Hoeft}
  et~al.}{2004}]{hoeft:04}
{Hoeft} M.,  {Br{\"u}ggen} M.,    {Yepes} G.,  2004, \mnras, 347, 389

\bibitem[\protect\citeauthoryear{{Hoeft}, {Br{\"u}ggen}, {Yepes} \&
  {Gottl{\"o}ber}}{{Hoeft} et~al.}{2006}]{hoeft:06}
{Hoeft} M.,  {Br{\"u}ggen} M.,  {Yepes} G.,    {Gottl{\"o}ber} S.,  2006, in
  preparation

\bibitem[\protect\citeauthoryear{{Jaffe}}{{Jaffe}}{1977}]{jaffe:77}
{Jaffe} W.~J.,  1977, \apj, 212, 1

\bibitem[\protect\citeauthoryear{{Jones} \& {Ellison}}{{Jones} \&
  {Ellison}}{1991}]{jones:91}
{Jones} F.~C.,  {Ellison} D.~C.,  1991, Space Science Reviews, 58, 259

\bibitem[\protect\citeauthoryear{{Kardashev}}{{Kardashev}}{1962}]{kardashev:62}
{Kardashev} N.~S.,  1962, Soviet Astronomy, 6, 317

\bibitem[\protect\citeauthoryear{{Kempner}, {Blanton}, {Clarke}, {En{\ss}lin},
  {Johnston-Hollitt} \& {Rudnick}}{{Kempner} et~al.}{2004}]{kempner:04}
{Kempner} J.~C.,  {Blanton} E.~L.,  {Clarke} T.~E.,  {En{\ss}lin} T.~A.,
  {Johnston-Hollitt} M.,    {Rudnick} L.,  2004, in {Reiprich} T.,  {Kempner}
  J.,   {Soker} N.,  eds, The Riddle of Cooling Flows in Galaxies and Clusters
  of galaxies {Conference Note: A Taxonomy of Extended Radio Sources in
  Clusters of Galaxies}.
pp 335--+

\bibitem[\protect\citeauthoryear{{Keshet}, {Waxman} \& {Loeb}}{{Keshet}
  et~al.}{2004}]{keshet:04}
{Keshet} U.,  {Waxman} E.,    {Loeb} A.,  2004, \apj, 617, 281

\bibitem[\protect\citeauthoryear{{Keshet}, {Waxman}, {Loeb}, {Springel} \&
  {Hernquist}}{{Keshet} et~al.}{2003}]{keshet:03}
{Keshet} U.,  {Waxman} E.,  {Loeb} A.,  {Springel} V.,    {Hernquist} L.,
  2003, \apj, 585, 128

\bibitem[\protect\citeauthoryear{{Levinson}}{{Levinson}}{1994}]{levinson:94}
{Levinson} A.,  1994, \apj, 426, 327

\bibitem[\protect\citeauthoryear{{Loken}, {Norman}, {Nelson}, {Burns}, {Bryan}
  \& {Motl}}{{Loken} et~al.}{2002}]{loken:02}
{Loken} C.,  {Norman} M.~L.,  {Nelson} E.,  {Burns} J.,  {Bryan} G.~L.,
  {Motl} P.,  2002, \apj, 579, 571

\bibitem[\protect\citeauthoryear{{Malkov} \& {O'C Drury}}{{Malkov} \& {O'C
  Drury}}{2001}]{malkov:01}
{Malkov} M.~A.,  {O'C Drury} L.,  2001, Reports of Progress in Physics, 64, 429

\bibitem[\protect\citeauthoryear{{Markevitch}, {Gonzalez}, {David},
  {Vikhlinin}, {Murray}, {Forman}, {Jones} \& {Tucker}}{{Markevitch}
  et~al.}{2002}]{markevitch:02}
{Markevitch} M.,  {Gonzalez} A.~H.,  {David} L.,  {Vikhlinin} A.,  {Murray} S.,
   {Forman} W.,  {Jones} C.,    {Tucker} W.,  2002, \apjl, 567, L27

\bibitem[\protect\citeauthoryear{{Medvedev}, {Fiore}, {Fonseca}, {Silva} \&
  {Mori}}{{Medvedev} et~al.}{2005}]{medvedev:05}
{Medvedev} M.~V.,  {Fiore} M.,  {Fonseca} R.~A.,  {Silva} L.~O.,    {Mori}
  W.~B.,  2005, \apjl, 618, L75

\bibitem[\protect\citeauthoryear{{Medvedev}, {Silva} \&
  {Kamionkowski}}{{Medvedev} et~al.}{2006}]{medvedev:06}
{Medvedev} M.~V.,  {Silva} L.~O.,    {Kamionkowski} M.,  2006, \apjl, 642, L1

\bibitem[\protect\citeauthoryear{{Miniati}, {Jones}, {Kang} \& {Ryu}}{{Miniati}
  et~al.}{2001}]{miniati:01}
{Miniati} F.,  {Jones} T.~W.,  {Kang} H.,    {Ryu} D.,  2001, \apj, 562, 233

\bibitem[\protect\citeauthoryear{{Miniati}, {Ryu}, {Kang}, {Jones}, {Cen} \&
  {Ostriker}}{{Miniati} et~al.}{2000}]{miniati:00}
{Miniati} F.,  {Ryu} D.,  {Kang} H.,  {Jones} T.~W.,  {Cen} R.,    {Ostriker}
  J.~P.,  2000, \apj, 542, 608

\bibitem[\protect\citeauthoryear{{Nagai} \& {Kravtsov}}{{Nagai} \&
  {Kravtsov}}{2003}]{nagai:03}
{Nagai} D.,  {Kravtsov} A.~V.,  2003, \apj, 587, 514

\bibitem[\protect\citeauthoryear{{Pfrommer}, {Springel}, {En{\ss}lin} \&
  {Jubelgas}}{{Pfrommer} et~al.}{2006}]{pfrommer:06}
{Pfrommer} C.,  {Springel} V.,  {En{\ss}lin} T.~A.,    {Jubelgas} M.,  2006,
  \mnras, 367, 113

\bibitem[\protect\citeauthoryear{{Rephaeli} \& {Gruber}}{{Rephaeli} \&
  {Gruber}}{2003}]{rephaeli:03}
{Rephaeli} Y.,  {Gruber} D.,  2003, \apj, 595, 137

\bibitem[\protect\citeauthoryear{{Roettiger}, {Burns} \& {Stone}}{{Roettiger}
  et~al.}{1999}]{roettiger:99}
{Roettiger} K.,  {Burns} J.~O.,    {Stone} J.~M.,  1999, \apj, 518, 603

\bibitem[\protect\citeauthoryear{{R{\"o}ttgering}, {Wieringa}, {Hunstead} \&
  {Ekers}}{{R{\"o}ttgering} et~al.}{1997}]{roettgering:97}
{R{\"o}ttgering} H.~J.~A.,  {Wieringa} M.~H.,  {Hunstead} R.~W.,    {Ekers}
  R.~D.,  1997, \mnras, 290, 577

\bibitem[\protect\citeauthoryear{{Rybicki} \& {Lightman}}{{Rybicki} \&
  {Lightman}}{1986}]{rybicki:86}
{Rybicki} G.~B.,  {Lightman} A.~P.,  1986, {Radiative Processes in
  Astrophysics}.
Radiative Processes in Astrophysics, by George B.~Rybicki, Alan P.~Lightman,
  pp.~400.~ISBN 0-471-82759-2.~Wiley-VCH , June 1986.

\bibitem[\protect\citeauthoryear{{Ryu}, {Kang}, {Hallman} \& {Jones}}{{Ryu}
  et~al.}{2003}]{ryu:03}
{Ryu} D.,  {Kang} H.,  {Hallman} E.,    {Jones} T.~W.,  2003, \apj, 593, 599

\bibitem[\protect\citeauthoryear{{Sarazin}}{{Sarazin}}{1999}]{sarazin:99}
{Sarazin} C.~L.,  1999, \apj, 520, 529

\bibitem[\protect\citeauthoryear{{Shibata}, {Honda}, {Ishida}, {Ohashi} \&
  {Yamashita}}{{Shibata} et~al.}{1999}]{shibata:99}
{Shibata} R.,  {Honda} H.,  {Ishida} M.,  {Ohashi} T.,    {Yamashita} K.,
  1999, \apj, 524, 603

\bibitem[\protect\citeauthoryear{{Slee}, {Roy}, {Murgia}, {Andernach} \&
  {Ehle}}{{Slee} et~al.}{2001}]{slee:01}
{Slee} O.~B.,  {Roy} A.~L.,  {Murgia} M.,  {Andernach} H.,    {Ehle} M.,  2001,
  \aj, 122, 1172

\bibitem[\protect\citeauthoryear{{Sun}, {Murray}, {Markevitch} \&
  {Vikhlinin}}{{Sun} et~al.}{2002}]{sun:02}
{Sun} M.,  {Murray} S.~S.,  {Markevitch} M.,    {Vikhlinin} A.,  2002, \apj,
  565, 867

\bibitem[\protect\citeauthoryear{{Vikhlinin} \& {Markevitch}}{{Vikhlinin} \&
  {Markevitch}}{2002}]{vikhlinin:02}
{Vikhlinin} A.~A.,  {Markevitch} M.~L.,  2002, Astronomy Letters, 28, 495

\bibitem[\protect\citeauthoryear{{Vink} \& {Laming}}{{Vink} \&
  {Laming}}{2003}]{vink:03}
{Vink} J.,  {Laming} J.~M.,  2003, \apj, 584, 758

\bibitem[\protect\citeauthoryear{{Wentzel}}{{Wentzel}}{1974}]{wentzel:74}
{Wentzel} D.~G.,  1974, \araa, 12, 71

\end{thebibliography}
\bibliographystyle{mn2e}

\clearpage

\begin{table}
\begin{center} 
\begin{tabular}{c|ccccc|cccc}
 & $z$ & $T$   & LLS  & $P_{1.4}$                                           & $\alpha$  & $\cal M$ & $B$      & $y_c$ & $\xi_B$ \\
 &     & [keV] &[Mpc] & $\left[10^{31}\:\frac{\rm erg}{\rm s \, Hz}\right]$ &           &          & [$\mu$G] & [Mpc] & $10^{-3}$ \\[.3ex]
\hline
\hline
Abell 115 & 0.1971   & 5   & 2.5  & 1.9  & 1.1  & 4.7   & 0.1  & 0.1 & 0.2 \\
Abell 2256 & 0.0594  & 7.8 & 1.1  & 3.6  & 1.2  & 3.3   & 0.3  & 0.4 & 1.1 \\
Abell 3667 & 0.055   & 8   & 2.0  & 41.0 & 1.1  & 4.7   & 0.2  & 0.3 & 0.5 \\
\hline
\end{tabular}
\caption{
Observed and derived parameters for our three model relics. For all computations we have assumed an electron density of $n_{\rm e} = 10^{-4}\:{\rm cm^{-3}}$. LLS indicates the longest linear extension of a relic. To determine the radio power of the relics we use $H_0 = 50 \:{\rm km \, s^{-1} \, Mpc^{-1}}$ in order to be consistent with computations in literature.
}
\label{tab-abell}
\end{center}
\end{table}

\clearpage

\clearpage

\clearpage

\clearpage

\clearpage

\clearpage

\clearpage

\clearpage

\end{document}